\documentclass{emulateapj}
\usepackage{amssymb,latexsym,amsmath,graphicx,url}
\pdfoutput=1
\usepackage{hyperref}
\usepackage{natbib}
\usepackage{longtable}
\slugcomment{Accepted for publication in The Astrophysical Journal}
\shorttitle{The Regulation of Cooling and Star Formation by AGN Feedback}
\shortauthors{Rafferty et al.}

\begin{document}

\setlength{\pdfpageheight}{\paperheight}
\setlength{\pdfpagewidth}{\paperwidth}

\title{The Regulation of Cooling and Star Formation in Luminous Galaxies by AGN Feedback and the Cooling-Time/Entropy Threshold for the Onset of Star Formation}

\author{D.~A.~Rafferty} 
\affil{Department of Astronomy and Astrophysics, The Pennsylvania State University, University Park, PA 16802}
\affil{Department of Physics and Astronomy, Ohio University, Athens, OH 45701}
\author{B.~R.~McNamara}
\affil{Department of Physics and Astronomy, University of Waterloo, Waterloo, ON N2L 2G1, Canada,}
\affil{Perimeter Institute for Theoretical Physics, 31 Caroline St. N., Waterloo, N2L 2Y5 Ontario, Canada,}
\affil{Harvard-Smithsonian Center for Astrophysics, Cambridge, MA 02138}
\and 
\author{P.~E.~J.~Nulsen}
\affil{Harvard-Smithsonian Center for Astrophysics, Cambridge, MA 02138}

\begin{abstract}
Using broadband optical imaging and \textit{Chandra} X-ray data for a sample of 46 cluster central dominant galaxies (CDGs), we investigate the connection between star formation, the intracluster medium (ICM), and the central active galactic nucleus (AGN). We report the discovery of a remarkably sharp threshold for the onset of star formation that occurs when the central cooling time of the hot atmosphere falls below $\sim 5 \times 10^8$ yr, or equivalently when the central entropy falls
below  $\sim 30$ keV cm$^2$.  In addition to this criterion, star formation in cooling flows also appears to require that the X-ray and galaxy centroids lie within $\sim 20$ kpc of each other, and that the jet (cavity) power is smaller than the X-ray cooling luminosity.  These three criteria, together with the high ratio of cooling time to AGN outburst (cavity) age across our sample, directly link the presence of star formation and AGN activity in CDGs to cooling instabilities in the intracluster plasma.  Our results provide compelling evidence that AGN feedback into the hot ICM is largely responsible for regulating cooling and star formation in the cores of clusters, leading to the significant growth of supermassive black holes in CDGs at late times.
\end{abstract}


\section{Introduction}
 X-ray observations of the intracluster medium (ICM) in a majority of clusters show stongly peaked central emission from gas with short cooling times \citep{pere98}. Such clusters were thought to harbor cooling flows \citep{fabi94}, in which gas cooling out of the ICM at the core of the cluster is replaced by gas at larger radii in a slow inward flow. Low-spectral-resolution X-ray observations predicted deposition rates of $100-1000\mbox{'s M$_{\odot}$ yr$^{-1}$}$ \citep[e.g.,][]{pere98}. These rates and the implied cold gas masses exceeded the inferred star formation rates \citep{mcna97} and the observed gas masses \citep{edge01} in the central galaxy by an order of magnitude. This apparent disagreement between the amount of gas seen cooling and the amount seen in traditional sinks persisted for more than two decades.
 
Recently, however, high-spectral resolution observations by the \textit{Chandra} and XMM-\textit{Newton} X-ray Observatories have shown that the cooling rates predicted by ROSAT are too large by a factor of ten \citep[e.g.,][]{davi01,blan03,pete03,kaas04}.  These new telescopes have shown that cooling-flow spectra lack the emission signatures expected from gas cooling below $\sim 1$ keV. Heating by the central active galactic nucleus (AGN) has emerged as the most promising means to prevent significant amounts of cooling from occurring \citep[for a review, see][]{mcna07}.
 
Some form of heating is also required in hierarchical CDM simulations to reproduce the high-luminosity end of the galaxy luminosity function. Recent simulations \citep[e.g.,][]{crot06,delu07} have found that strong feedback at the cores of the largest halos is necessary to prevent over-cooling and the formation of much more massive galaxies than are observed. The local galaxy luminosity function shows an exponential cutoff at the highest luminosities \citep[e.g.,][]{bens03,bowe06} that is reproduced well by models of hierarchical formation that include AGN feedback. 

However, some net cooling is probably occurring, as evidence of active star formation is observed at the cores of many cooling flows \citep[e.g.][]{john87,roma87,alle95,smit97,hick05,dona07,odea08}, and it has been known for some time that indicators of star formation in the centrally dominant galaxy (CDG), such as spatially extended excess blue emission and optical line emission, correlate, albeit roughly, with the properties of the cooling flow \citep[e.g.,][]{heck81,hu85,mcna89,card95}. Additionally, CDGs with strong optical line emission are preferentially found near the cores of cooling flows, implying a physical connection between star formation and the cooling ICM \citep{craw99,edwa07}.

However, only recently, with the advent of \textit{Chandra}, has it become possible to derive the X-ray properties on the same scales as the optical properties to determine more precisely the relationship between cooling and star formation. In this paper, we use optical and X-ray data of similar spatial resolution for 46 CDGs to examine the connection between the ICM properties, such as the central cooling time and the AGN's feedback power, and the presence of star formation in the central galaxy. We adopt $H_0 = 70$~km~s$^{-1}$ Mpc$^{-1}$, $\Omega_{\Lambda} = 0.7$, and $\Omega_{\rm{M}} = 0.3$ throughout this paper.

\section{The Sample}
The sample comprises 39 CDGs for which we have obtained broadband optical data, plus an additional 7 systems for which we have taken optical data from the literature. The sample was constructed to test whether the presence of active star formation in the CDG is related to the properties of the cooling flow. To that end, the objects in the sample were chosen from the \textit{Chandra} Data Archive to include both cooling flows and noncooling flows, with a wide range of central cooling times, from $t_{\rm cool} \lesssim 10^8$ yr to $t_{\rm cool} \gtrsim 10^{10}$ yr. Additional constraints on the sample were that the available X-ray images have sufficient counts to derive density and temperature profiles and that the systems be visible to the optical observatory during the alloted observing windows. Table \ref{T:sample} lists the general properties of the objects in the sample.

\def\arraystretch{1.2}
\tabletypesize{\scriptsize}
\begin{deluxetable*}{llccccccc}
\tablewidth{0pt}
\tablecaption{Sample Properties. \label{T:sample}}
\tablehead{
\colhead{} & \colhead{} & \multicolumn{2}{c}{X-ray Core (J2000)} & \colhead{} & \colhead{} & \multicolumn{2}{c}{CDG Core (J2000)} & \colhead{$\Delta r$\tablenotemark{b}} \\
\cline{3-4} \cline{7-8}
\colhead{System\phn\phn\phn} & \colhead{$z$} & \colhead{$\alpha$ ($\degr$)} & \colhead{$\delta$ ($\degr$)} &  \colhead{CDG Name} & \colhead{$M_{K}$\tablenotemark{a}}  & \colhead{$\alpha$ ($\degr$)} & \colhead{$\delta$ ($\degr$)} & \colhead{(kpc)} }
\startdata                                                                                         
   A85               & 0.055 &  10.4593 &  -9.3022 & PGC 002501 			& $-26.72\pm 0.04$&  10.4603 	&  -9.3031 &    5.2 		\\
   3C 28             & 0.195 &  13.9599 &  26.4098 & PGC 138263 			& $-26.08\pm 0.18$&  13.9609 	&  26.4105 &   12.9	\\
   A133              & 0.057 &  15.6744 & -21.8804 & ESO 541-013 			& $-26.37\pm 0.06$&  15.6739 & -21.8820 &    6.5 	\\
   A223              & 0.207 &  24.4828 & -12.8198 & 2MASX J01375602-1249106& $-26.36\pm 0.16$&  24.4833 & -12.8195 & $<  6.8$\\
   A262              & 0.016 &  28.1922 &  36.1528 & NGC 708 				& $-25.66\pm 0.03$&  28.1936 				&  36.1520 &    1.7 	\\
   A383              & 0.187 &  42.0139 &  -3.5291 & PGC 145057 			& $-26.89\pm 0.12$&  42.0141 				&  -3.5291 & $<  6.3$\\
   AWM 7             & 0.017 &  43.6145 &  41.5797 & NGC 1129  				& $-26.09\pm 0.03$&  43.6139 				&  41.5796 & $<  0.7$\\
   Perseus           & 0.018 &  49.9507 &  41.5118 & NGC 1275 				& $-26.29\pm 0.04$&  49.9510 				&  41.5116 & $<  0.7$\\
   2A 0335+096       & 0.035 &  54.6700 &   9.9660 & PGC 013424 			& $-26.18\pm 0.05$&  54.6691				&   9.9701 &   10.6 	\\
   A478              & 0.088 &  63.3551 &  10.4653 & PGC 014685 			& $-26.67\pm 0.07$&  63.3553 				&  10.4652 & $<  3.3$\\
   A496              & 0.033 &  68.4076 & -13.2619 & PGC 015524  			& $-26.28\pm 0.04$&  68.4077 			& -13.2619 & $<  1.3$\\
   A520              & 0.199 &  73.5422* &   2.9248* & PGC 1240180  		& $-25.68\pm 0.27$&  73.5160 			&   2.8923 &  494.4 	\\
   MS 0735.6+7421    & 0.216 & 115.4351 &  74.2440 & PGC 2760958 			& $-26.37\pm 0.17$& 115.4361 			&  74.2438 & $<  7.0$\\
   PKS 0745-191      & 0.103 & 116.8806 & -19.2947 & PGC 021813 			& $-26.82\pm 0.09$& \nodata 				& \nodata  & \nodata \\
   Hydra A           & 0.055 & 139.5238 & -12.0953 & PGC 026269 			& $-25.91\pm 0.06$& \nodata 				& \nodata  & \nodata \\
   Zw 3146           & 0.290 & 155.9152 &   4.1863 & 2MASX J10233960+0411116& $-27.67\pm 0.14$& \nodata & \nodata & \nodata  \\
   A1068	     	 & 0.138 & 160.1853 &  39.9532 & PGC 093944 			& $-26.70\pm 0.08$& \nodata 					& \nodata  & \nodata \\
   A1361             & 0.117 & 175.9150 &  46.3562 & PGC 093947 			& $-26.29\pm 0.09$& 175.9150 				&  46.3556 &    4.6 	\\
   A1413             & 0.143 & 178.8249 &  23.4052 & PGC 037477 			& $-27.31\pm 0.08$& 178.8250 				&  23.4049 & $<  5.0$\\
   M87               & 0.004 & 187.7057 &  12.3913 & M87 					& $-25.47\pm 0.02$& 187.7059 					&  12.3912 & $<  0.2$\\
   HCG 62            & 0.014 & 193.2740 &  -9.2036 & NGC 4778 				& $-25.26\pm 0.03$& 193.2738 				&  -9.2040 & $<  0.6$\\
   A1650             & 0.084 & 194.6727 &  -1.7617 & PGC 1110773 			& $-25.98\pm 0.10$& 194.6730 			&  -1.7614 & $<  3.2$\\
   Coma              & 0.023 & 194.8984* &  27.9591* & NGC 4874 			& $-26.08\pm 0.03$& 194.8988 				&  27.9593 &   85.8 	\\
   A1795             & 0.063 & 207.2196 &  26.5913 & PGC 049005 			& $-26.49\pm 0.08$& 207.2188 				&  26.5929 &    7.5 	\\
   A1835             & 0.253 & 210.2581 &   2.8789 & 2MASX J14010204+0252423& $-27.36\pm 0.14$& \nodata 	& \nodata  & \nodata \\
   A1991             & 0.059 & 223.6314 &  18.6445 & NGC 5778 				& $-26.18\pm 0.08$& 223.6313 				&  18.6424 &    8.8 	\\
   MS 1455.0+2232    & 0.258 & 224.3128 &  22.3424 & PGC 1668167 			& $-27.14\pm 0.14$& 224.3130 			&  22.3429 & $<  8.0$\\
   RXC J1504.1-0248  & 0.215 & 226.0310 &  -2.8043 & PGC 126345  			& $-26.58\pm 0.18$& 226.0313 			&  -2.8045 & $<  7.0$\\
   A2029             & 0.077 & 227.7337 &   5.7449 & PGC 054167 			& $-27.41\pm 0.05$& \nodata 				& \nodata  & \nodata \\
   A2052             & 0.035 & 229.1851 &   7.0215 & UGC 09799 				& $-26.30\pm 0.06$& \nodata 				& \nodata  & \nodata \\
   A2065             & 0.073 & 230.6224 &  27.7052 & PGC 054888 			&\nodata& 230.6215 				&  27.7077 &   13.2 	\\
   RX J1532.8+3021   & 0.345 & 233.2242 &  30.3497 & PGC 1900245  			& \nodata& 233.2241 			&  30.3499 & $<  9.8$\\
   A2218             & 0.176 & 248.9613 &  66.2110 & PGC 140648 			& \nodata& 248.9553 				&  66.2124 &   30.1 	\\
   Hercules A        & 0.154 & 252.7841 &   4.9924 & PGC 059117 			& $-26.45\pm 0.11$& 252.7840 				&   4.9927 & $<  5.3$\\
   A2244             & 0.097 & 255.6775 &  34.0606 & PGC 140689 			& $-26.97\pm 0.09$& 255.6772 				&  34.0604 & $<  3.6$\\
   NGC 6338          & 0.027 & 258.8453 &  57.4113 & NGC 6338 				& $-25.90\pm 0.03$& 258.8456 				&  57.4114 & $<  1.1$\\
   RX J1720.2+2637   & 0.164 & 260.0413 &  26.6250 & PGC 1782937 			& $-26.68\pm 0.11$& 260.0419 			&  26.6256 &    8.5 	\\
   MACS J1720.2+3536 & 0.391 & 260.0699 &  35.6075 & \nodata 				& \nodata & 260.0699 				&  35.6074 & $< 10.6$\\
   A2261             & 0.224 & 260.6132 &  32.1327 & PGC 1981854 			& $-27.35\pm 0.09$& 260.6133 			&  32.1326 & $<  7.2$\\
   A2319             & 0.056 & 290.3044* &  43.9366* & PGC 063099 			& $-26.75\pm 0.06$& 290.2918 			&  43.9456 &   50.1 	\\
   A2390             & 0.228 & 328.4034 &  17.6955 & PGC 140982 			& $-27.05\pm 0.17$& 328.4035 				&  17.6957 & $<  7.3$\\
   A2409             & 0.148 & 330.2194 &  20.9743 & PGC 093957 			& $-27.03\pm 0.10$& 330.2191 				&  20.9693 &   46.6 	\\
   A2597             & 0.085 & 351.3322 & -12.1239 & PGC 071390 			& $-25.63\pm 0.11$& 351.3322 				& -12.1242 & $<  3.2$\\
   A2626             & 0.055 & 354.1271 &  21.1471 & IC 5338 				& $-26.37\pm 0.06$& 354.1270 				&  21.1464 &    2.6 	\\
   A2657             & 0.040 & 356.2393 &   9.1919 & PGC 072297 			& $-25.48\pm 0.07$& 356.2393 				&   9.1932 &    3.5 	\\
   A2670             & 0.076 & 358.5571 & -10.4189 & PGC 072804 			& $-26.82\pm 0.09$& 358.5570 				& -10.4190 & $<  2.9$\\
\enddata
\tablecomments{The X-ray and CDG core positions are from this work; see Sections \ref{S:X-ray_analysis} and \ref{S:profiles}. Systems that lack optical CDG positions were not observed in this study; optical data for these systems was taken from the literature (see references given in Table \ref{T:gradients}). Uncertain core positions (due to the lack of a cooling core or to the presence of confusing substructure) are marked with an asterisk.}
\tablenotetext{a}{Total $K$-band magnitudes from the 2MASS catalog.}
\tablenotetext{b}{Projected radius from the X-ray core to the CDG's core. Separations of less than 2 arcsec are treated as upper limits.}
\end{deluxetable*}

\section{Observations and Data Reduction}
\subsection{Optical Data}
The primary signatures of star formation are due to the presence of hot, massive stars in the star-forming population. This population is mixed with the generally old stellar population of the underlying CDG, as well as with cold gas and dust. These massive stars emit most of their energy in the ultraviolet and blue parts of the spectrum, unlike the less massive but much longer-lived stars that account for most of the CDG's emission at longer wavelengths. The ultraviolet emission in turn can ionize nearby cool gas, resulting in optical line emission such as H$\alpha$ emission \citep[e.g.,][]{heck89,voit97,hatc07}. Additionally, the energetic photons from the massive stars heat surrounding dust, which re-emits the light at far infrared wavelengths \citep{odea08}. In this study, we use the excess blue emission (above that expected from the old background population) as an indicator of recent star formation.

We use broadband imaging at short and long wavelengths to search for the excess blue emission in the cores of the CDGs. By comparing the short-wavelength surface-brightness profile, which is sensitive to young, hot stars, with the long-wavelength profile, which traces the old background population, we can detect the presence of star formation. Elliptical galaxies generally show no signs of recent star formation, and have color profiles that become bluer with increasing radius \citep[e.g.,][]{vade88,fran89,pele90,goud94}. This effect is thought to be generally due to a decreasing stellar metalicity with increasing radius \citep[e.g.,][]{caro93,koba99,tamu00,ferr05}. Therefore, a color profile that becomes increasingly blue towards the center is unusual (i.e.\ the CDG has excess blue emission), and is indicative of recent star formation in that region. We use this property to identify star formation in our sample. 

The excess blue emission, while generally indicative of star formation, can be the result of other sources, such as the scattered light from an AGN or a low-metalicity stellar population.  However, because AGNs are point sources, their emission should not be spatially extended on scales of tens of kpc, as the excess blue emission is often observed to be in CDGs. Scattered light can make the AGN emission appear more extended, but studies of the nearby cooling flows A1795 and A2597 have ruled out significant polarization in the excess blue light \citep[e.g.,][]{mcna96a,mcna99}. As to metalicity effects, most modeling of spectra and broadband colors of CDGs has found that systems with large color excesses are better modeled by emission from massive O and B stars, rather than old, low-metalicity stellar populations \citep[e.g.,][]{alle95,card95,smit97,card98,craw99}. Additionally, Hubble Space Telescope studies of two CDGs with prominent blue excesses, A1795 and A2597, have resolved the blue emission into knots and demonstrated that it is most likely due to recent star formation \citep{koek99,koek02,odea04}.

Since our goal is to search for connections between star formation and the cooling flow, we used data that are sensitive to the presence of active star formation. Such star formation is best detected at short wavelengths due to the rising spectral energy distribution (SED) of hot stars in the ultraviolet and the falling SED of the background galaxy population. Therefore, we chose to observe in the \textit{U} band (with a central wavelength of 3582 \AA).  This choice minimizes contamination from redshifted [OII]$\lambda$3227 emission lines expected to be present in many of the systems.  We also need long-wavelength observations to trace the old stellar population; however, at long wavelengths, contamination from H$\alpha$ and H$\beta$ lines can be a problem; therefore, we chose to observe in both \textit{R} and \textit{I} filters (with central wavelengths of 6513 \AA\ and 8204 \AA, respectively).  Colors obtained from both filters give us a means of verifying that any anomalous blue emission is due to star formation. If the $U-R$ and $U-I$ colors are both anomalously blue, it is likely to be due to star formation, and not due to contamination from emission lines (at least in the red images).

Optical data were obtained during five separate runs. Three runs, totaling 11 nights, were done using the 2.4-m telescope at the MDM observatory on Kitt Peak, Arizona between the dates of March 9-12, 2005, September 26-29, 2005, and May 22-24, 2006. An additional run of four nights was performed at the KPNO Mayall 4-m telescope between October 5-8, 2005. Lastly, a 2-night run was performed at the WIYN 3.5-m telescope over January 22-23, 2006. See Table~\ref{T:observations} for details of the optical observations. Broadband imaging was done at MDM using the Echelle CCD with a $9.4 \times 9.4$ arcmin field of view and a scale of 0.28 arcsec pixel$^{-1}$. At the KPNO 4-m telescope, imaging was done using the Mosaic CCD array of eight CCDs with a $36 \times 36$ arcmin field of view and a scale of 0.26 arcsec pixel$^{-1}$. At the WIYN 3.5-m telescope, imaging was done using the Mini-Mosaic CCD array of two CCDs, with a $9.6 \times 9.6$ arcmin field of view and a scale of 0.14 arcsec pixel$^{-1}$. 

Harris $U, R,$ and $I$ filters were used during all runs. All objects were imaged in the $U$ band and in either $R$ or $I$ band (most objects were imaged in both $R$ and $I$ bands). Exposure times were typically $2 \times 600$ s in $U,$ $2 \times 150$ s in $I,$ and $2 \times 225$ s in $R$; however, in some cases (e.g., for distant systems), longer total exposure times were required to achieve sufficient signal to noise. Multiple frames were taken to allow for easy removal of cosmic-ray events. The frames were dithered by $\sim 30 - 60$ arcsec between exposures to allow for the removal of artifacts such as bad columns and CCD gaps and to improve flat fielding. Conditions varied during the runs, with photometric conditions for approximately two-thirds of the total time. 

\begin{deluxetable*}{lccccccc}
\tablewidth{0pt}
\tablecaption{X-ray and Optical Observations. \label{T:observations}}
\tablehead{
	\colhead{} & \multicolumn{2}{c}{X-ray Observations} & \colhead{} & \multicolumn{4}{c}{Optical Observations} \\
	\cline{2-3} \cline{5-8} 
	\colhead{} & \colhead{} & \colhead{Exp. Time\tablenotemark{a}} & \colhead{} & \colhead{} & \multicolumn{3}{c}{Total Exp. Time (s)}  \\
	\colhead{System\phn} & \colhead{OBSID} & \colhead{(ks)} & \colhead{} & \colhead{Telescope} & \colhead{$U$} & \colhead{$R$} & \colhead{$I$} }
\startdata
A85		& 904  & 37.5 &  & 2.4 m & 1200		& 300		& 450			\\
3C28		& 3233 & 48.2 &  & 2.4 m & 1800		& 300		& 450			\\
A133		& 2203 & 30.9 &  & 4 m	 & 2000		& 200		& 800			\\
A223		& 4967 & 42.2 &  & 4 m	 & 2200		& 1600		& 300			\\
A262		& 2215 & 26.7 &  & 4 m	 & 1200		& \nodata	& 1000			\\
A383		& 2321 & 16.8 &  & 2.4 m & 1800		& 450		& 225			\\
AWM 7		& 908  & 47.9 &  & 2.4 m & 600 		& \nodata 	& 450 		 	\\
Perseus		& 4947 & 29.6 &  & 4 m	 & 3600		& \nodata	& 500			\\
2A 0335+096	&  919 & 16.1 &  & 2.4 m & 1200		& 300		& 450			\\
A478		& 1669 & 41.0 &  & 4 m	 & 1800		& 700		& \nodata		\\
A496		& 4976 & 58.0 &  & 2.4 m & 1200		& 300		& 450			\\
A520		& 4215 & 54.2 &  & 4 m	 & 3200		& 300		& 1200			\\
MS 0735.6+7421	& 4197 & 39.9 &  & 3.5 m & 2100		& 3600 		& 2400			\\
PKS 0745-191	& 2427 & 17.4 &  & \nodata & \nodata 	& \nodata 	& \nodata 	  \\
Hydra A		& 4970 & 98.8 &  & \nodata & \nodata 	& \nodata 	& \nodata 	  \\
Zw 3146		&  909 & 41.5 &  & \nodata & \nodata 	& \nodata 	& \nodata 	 \\
A1068		& 1652 & 25.6 &  & \nodata & \nodata 	& \nodata 	& \nodata 	 \\
A1361		& 3369 &  3.0 &  & 2.4 m & 1800		& 450		& 225			\\
A1413		& 5003 & 65.8 &  & 2.4 m & 2700		& 900		& 450			\\
M87		& 3717 & 15.1 &  & 2.4 m & 600 		& 200		& 345 		 \\
HCG 62		& 921  & 47.5 &  & 2.4 m & 300		& 150		& 150			\\
A1650		& 4178 & 26.6 &  & 2.4 m & 300		& 150		& 150			\\
Coma		& 1086 &  9.3 &  & 2.4 m & 900		& 300		& \nodata		\\
A1795		& 3666 & 14.2 &  & 2.4 m & 1800		& 450		& 225			\\
A1835		&  496 & 10.3 &  & \nodata & \nodata 	& \nodata 	& \nodata 	\\
A1991		& 3193 & 35.8 &  & 2.4 m & 1800		& 450		& 225			\\
MS 1455.0+2232	& 4192 & 83.2 &  & 2.4 m & 1800		& 450		& 225			\\
RXC J1504.1-0248& 5793 & 33.2 &  & 2.4 m & 1800		& 600		& 900			\\
A2029		& 4977 & 77.3 &  & \nodata & \nodata 	& \nodata 	& \nodata 	\\
A2052		& 890  & 36.1 &  & \nodata & \nodata 	& \nodata 	& \nodata 	 \\
A2065		& 3182 & 26.0 &  & 2.4 m & 1800		& 450		& 225			\\
RX J1532.8+3021	& 1649 &  8.8 &  & 2.4 m & 3600		& 1200		& 900			\\
A2218		& 1666 & 34.2 &  & 2.4 m & 1800		& 400		& 600		\\
Hercules A	& 1625 & 12.5 &  & 4 m	 & 2400		& 1200	 	& \nodata \\
A2244		& 4179 & 55.7 &  & 2.4 m & 1500		& 450		& 600			\\
NGC 6338	& 4194 & 44.0 &  & 2.4 m & 1800		& 450		& 225			\\
RX J1720.2+2637	& 4361 & 22.0 &  & 2.4 m & 1200		& 300		& 300			\\
		&      &      &  & 4 m	 & 2400		& \nodata 	& 600	 	\\
MACS J1720.2+3536& 6107 & 29.7 &  & 2.4 m & 1800		& 450		& 225			\\
A2261		& 5007 & 21.7 &  & 2.4 m & 3600		& 1050		& 1800		\\
A2319		& 3231 & 14.1 &  & 2.4 m & 1200		& \nodata	& 600	\\
A2390		& 4193 & 83.3 &  & 2.4 m & 1200		& 300		& 450			\\
		&      &      &  & 4 m	 & 2400		& 1100		& 1000		\\
A2409		& 3247 &  9.7 &  & 2.4 m & 1800		& 450		& 675			\\
A2597		&  922 & 11.3 &  & 2.4 m & 600		& 150		& 450		\\
A2626		& 3192 & 23.5 &  & 2.4 m & 1200		& 300		& 450			\\
A2657		& 4941 & 15.6 &  & 4 m	 & 1200		& 600		& 400			\\
A2670		& 4959 & 33.5 &  & 4 m	 & 1200		& \nodata	& 600		\\
\enddata
\tablenotetext{a}{Exposure time after cleaning for background flares.}
\end{deluxetable*}

Optical reduction and analysis was performed using IRAF (the Image Reduction and Analysis Facility\footnote{See \url{http://iraf.noao.edu/}.}), version 12.2.1, and custom procedures written in IDL.\footnote{See \url{http://www.ittvis.com/idl/}.} 

\subsubsection{Standard Reductions}\label{S:standard_reductions}
The bias level and its row-to-row variation were modeled and subtracted in all frames by fitting a high-order function to the overscan region of each chip.  The remaining bias structure was removed with a bias frame constructed by averaging 30-40 individual frames taken each night in the evening and morning. Twilight flats were then used to remove differences in pixel sensitivities across the chip. Flats were made for each night by averaging together at least 3 dithered frames taken during the evening and morning twilight with exposure levels of $\sim 1/2$ the saturation level of the CCDs.

The MDM Echelle CCD was found to have significant dark current ($\sim 0.1 - 0.2$ counts s$^{-1}$) during the March and September 2005 runs (the dark current during the May 2006 run was much lower), which resulted in large sky gradients in the long-exposure $U$-band frames. Additionally, a small but significant light contamination (probably due to light from the red light-emitting diodes on the instrument mounting) was present in the $R$- and $I$-band dark frames. We corrected for the dark current and light leak by creating a master dark frame in each band by averaging 4-5 individual frames together and subtracting a scaled version of this master frame from each object frame. Each individual dark frame had a dark time typical of the longest exposure time used during observation (e.g., 900 s for $U$). Unfortunately, the dark current was found to vary both in magnitude (by a factor of two) and in spatial structure over the course of a single night. This variation made it impossible to use a simple scaling by dark time. Instead, the best scaling was determined using a $\chi^2$ minimization routine. 

The best dark scaling for each object frame was determined by minimizing the residuals between the modes of the pixel values in multiple regions of the CCD after dark subtraction. This method effectively scales the dark frame to produce the flattest sky across the CCD. However, for this method to be effective, sky variations due to pixel-to-pixel differences in sensitivity must first be removed by flat fielding. Therefore, the dark images were also flat fielded before fitting and subtraction.  The fitting was done in IDL using the MPFIT package.\footnote{See \url{http://cow.physics.wisc.edu/~craigm/idl/fitting/}.}  Typically 9 object-free sub-regions of the chip in which the dark structure was most pronounced were used for fitting. If the fitted scaling in any given region differed by more than 3 sigma from the adopted mean it was rejected. The mean scaling across all the remaining regions was then used for dark subtraction.  Remaining sky gradients were typically on the order of 2-3\% of sky across the CCD, but they are generally much smaller ($\sim 0.5$\%) in the region of the BCG.

Fringing due to night sky lines was found to be significant in the $I$-band images. Therefore, a fringe frame for each night was constructed and subtracted from the $I$-band object frames. The fringe frame was made by averaging together all the $I$-band object frames of a given night, after masking of all objects (using the \textit{objmasks} task in IRAF). The sky level and any large-scale gradient were removed from the fringe frame using a median filter over large blocks of pixels to filter out small-scale variations. The final fringe frame was then scaled and subtracted from all the $I$-band object frames using the \textit{rmfringe} tool in IRAF.

In the KPNO data, an additional additive feature was present in $I$- and $U$-band images due to scattered light off of the prime-focus corrector (this feature is not present in the $R$-band images due to the anti-reflective coating applied to the corrector). A template image of this ``pupil ghost''  was made using the \textit{mscpupil} task in IRAF. This task isolates the pupil contribution to an image by fitting a spline function to the background of a master image and subtracting it off. The master image was made by averaging all flat-corrected images in a given filter over the entire run, after the masking of all objects using the \textit{objmask} task in IRAF, and scaling by the mode of pixel values in object-free regions of the image. The resulting pupil template image was then scaled and subtracted off of each image interactively.

Once the additive features were subtracted from each frame, a dark-sky flat was made by again averaging all frames in a given filter over each night. Generally, 10-20 frames were required to produce good dark-sky flats. When an insufficient number of frames existed for a given night (due, e.g., to poor weather), a dark-sky flat from a neighboring night was used. Once again, all objects were masked before averaging. The sky flats were then median smoothed with a $129 \times 129$-pixel box filter to reduce the noise but preserve large-scale structure.

Finally, a world coordinate system (WCS) solution was derived for each frame using the WCS mapping tools in IRAF. The WCS was accurately aligned and scaled to each frame using fits to USNO A2 catalog positions of typically 100--200 stars. After this adjustment, the WCS was found to be accurate to $\pm$0.3 arcsec or better.

\subsubsection{Sky subtraction}
Accurate determination of the sky surface brightness is essential for tracing galaxy profiles to large radius, where the galaxy's surface brightness is often just a few percent of the sky. Our study is concerned mainly with the profiles in the centers of the galaxies, where their emission dominates over the sky. However, we have nonetheless attempted to determine the sky level as accurately as possible. To this end, we have adopted the method of sky determination used in \citet{mcna92}, in which counts in source-free regions of the sky are assumed to be dominated by Poisson noise.  The distribution of counts in each region should then be well represented by a gaussian whose width is an estimate of the statistical error in the sky level. Therefore, to determine the sky level, we extract the counts in a number of source-free regions, construct a histogram of sky values, and fit a gaussian to the histogram. This method of fitting the distribution of sky values, in contrast to using a simple mode or mean, allows one to use the distribution's deviation from a normal distribution to identify the presence of contaminating sources.  

We use the mean of the best-fit gaussian as the modal sky value for a given region. We used 7-9 sky regions with areas of at least 1000 pixels each, distributed over the chip in the region of the CDG (but far enough away to minimize contamination), and eliminated regions whose sky modes were more than $2\sigma$ from the adopted mean. Comparison with sky values determined by fitting a plane to the entire CCD (median filtered over large blocks) showed typical differences in sky values between the two methods of $\lesssim 1-2$\%. However, our method of using smaller, source-free regions near the CDG should not be as sensitive as the plane-fitting method to residual large-scale gradients that may remain due to the dark current problems discussed in Section \ref{S:standard_reductions} and that may bias the derived sky level in the region of the CDG.

We also use the difference between sky modes in the multiple regions to estimate the systematic error in our sky estimate. As in \citet{mcna92}, we use the maximum difference between adopted sky modes ($\Delta S_{\rm{max}} = S_{\rm{max}} - S_{\rm{min}}$) to define the systematic error in sky levels as $\Delta S_{\rm{sys}} = \Delta S_{\rm{max}}/2$. When computing total errors in colors, systematic errors were added in quadrature with the statistical errors. 

\subsubsection{Image stacking}\label{S:stacking}
After sky subtraction and before stacking, the pixel values were changed to units of electrons per second  by multiplying by the gain and dividing by the exposure time. Next, the individual frames in each band were scaled to remove differences in intensity due to changes in airmass between exposures. The IRAF task \textit{mscimatch} was used to perform this scaling. \textit{Mscimatch} uses the measured intensities of unsaturated stars to determine the scalings; typically, 50-200 stars per frame were used. Since the frames for a given object were always taken within a short time of one another, scalings were typically $\lesssim 1$\%. We note that \textit{mscimatch} was not used to determine zero-point offsets, as this step was done during sky subtraction.

In order to stack the frames into a final master image, the frames must be tangent-plane projected onto the same pixel grid. To minimize artifacts due to projection, cosmic rays were identified and removed using the \textit{craverage} and \textit{fixpix} tasks in IRAF. \textit{Mscimage} was used to perform the projection using a sinc interpolant. Next, in the rare cases where there were large point spread function (PSF) variations between frames, PSF matching was done to remove these variations. Finally, the frames were stacked using a median filter (or an average when only 2 frames were used) and a $\sigma$-clipping routine with the \textit{mscstack} task in IRAF.  

\subsubsection{Calibration}
For the purposes of this paper, in which we compare colors measured between different radii of the same source, photometric calibration is unnecessary and was not done. While it is possible that we may miss significant active star formation if it is distributed smoothly across the galaxy (resulting in possibly negative color gradients), calibrated colors for similar systems \citep[e.g.,][]{mcna92} are consistent with there being little or no active star formation. Therefore, we consider it unlikely that significant star formation is present in any of the systems in our sample with negative color gradients.

Usually, because our sources are scattered across the sky, corrections must be made to remove extinction due to dust in our own galaxy, which varies as a function of position on the sky. However, we are measuring radial gradients and hence comparing colors between two radii in the same source. Since the Galactic extinction does not vary significantly over the typical angular scale of our galaxies ($r \lesssim 2$ arcmin), no correction for Galactic extinction is necessary. 

\subsection{X-ray Data}
All systems were observed with the \textit{Chandra} ACIS detector in imaging mode and the data were obtained from the \textit{Chandra} Data Archive. Details of the observations are given in Table \ref{T:observations}.

The Chandra data were reprocessed with CIAO 3.3 using CALDB 3.2.0 and were corrected for known time-dependent gain and charge transfer inefficiency problems.  Blank-sky background files, normalized to the count rate of the source image in the $10-12$ keV band, were used for background subtraction.\footnote{See \url{http://asc.harvard.edu/contrib/maxim/acisbg/}.} 

\section{Data Analysis and Results}\label{S:Analysis}
\subsection{Derivation of ICM Properties}\label{S:X-ray_analysis}
X-ray spectra were extracted in elliptical annuli with $\sim 3000$ counts and were centered on the centroid of the cluster emission with eccentricity and position angle set to the average values of the cluster isophotes. In systems with clear cooling cusps, the X-ray centroid was fixed to the position of the core, determined by fitting a gaussian to the surface brightness distribution in that region. In systems without cooling cores, the X-ray centroid was determined by fitting an ellipse to the image at an intermediate radius, usually at radii beyond any substructure. However, in some of these systems (e.g., A520), due to the diffuse nature of the cluster emission or to substructure, it was very difficult to identify the cluster center. Therefore, the centroid positions for these systems (noted in Table \ref{T:sample}) likely have large uncertainties. Spectra were extracted using \textit{dmextract} in CIAO, and weighted response files were made using the CIAO tools \textit{mkwarf} and \textit{mkacisrmf} or \textit{mkrmf} (\textit{mkacisrmf} was used for all observations taken at the $-120$ C focal plane temperature;  \textit{mkrmf} was used for all other observations).

Gas temperatures and densities were found by deprojecting the spectra with a single-temperature plasma model (MEKAL) with a foreground absorption model (WABS) using the PROJCT mixing model in XSPEC 11.3.2, between the energies of 0.5 keV and 7.0 keV. The redshift was fixed to the value given in Table \ref{T:sample}, and the foreground hydrogen column density was fixed to the Galactic value of \citet{dick90}, except in the cases of 2A 0335+096 and A478, when a significantly different value was preferred by the fit. In these two cases, the column density in each annulus was allowed to vary. The density was then calculated from the normalization of the MEKAL component, assuming $n_{\rm{e}}=1.2n_{\rm{H}}$ (for a fully ionized gas with hydrogen and helium mass fractions of $X=0.7$ and $Y=0.28$), as:
\begin{equation}
 n_{\rm e}=\sqrt{\frac{1.2\times10^{14} (4 \pi D_L^2) \times {\rm norm}}{(1+z)^2 V}},
\end{equation}
where $n_{\rm{e}}$ has units of cm$^{-3}$, the luminosity distance ($D_L$) has units of cm, and the volume of the shell ($V$) has units of cm$^{3}$. 

We derived the cooling times using the deprojected densities and temperatures found above and the cooling curves of  \citet{smit01}, calculated using APEC\footnote{See \url{http://cxc.harvard.edu/atomdb/}.} \citep[the results do not change significantly if we use the curves of][]{boeh89}. The pressure in each annulus was calculated as $p=nkT,$ where we have assumed an ideal gas and $n \approx 2n_{\rm{e}}$. Lastly, the entropy is defined as in \citet{lloy00}:
\begin{equation}
 S=kTn_{\rm e}^{-2/3},
\end{equation}
and has units of keV cm$^2$. 

A primary objective of this study is to determine how the ICM properties relate to the optical colors. In particular, we are interested in properties at the core of the cluster, where the cooling times are the shortest. However, the systems in our sample span a wide range of redshift and core surface brightness, resulting in a wide range in the physical size of the innermost region (which we required to contain $\sim 3000$ counts). 
\begin{deluxetable*}{lcccccc}
\tablewidth{0pt}
\tablecaption{Central ICM Properties. \label{T:ICM_central}}
\tablehead{
	\colhead{} & \multicolumn{2}{c}{$r$\tablenotemark{a}} & \colhead{$n_e$} & \colhead{$kT$} & \colhead{Entropy} & \colhead{$t_{\rm{cool}}$} \\
	\cline{2-3}
	\colhead{System} & \colhead{(arcsec)} & \colhead{(kpc)} & \colhead{(cm$^{-3}$)} & \colhead{(keV)} & \colhead{(keV cm$^2$)} & \colhead{(10$^8$ yr)} }
\startdata
   A85               &   2.4 &   2.6 & $0.107_{-0.008}^{+0.009}$ & $ 2.06_{-0.17}^{+0.15}$ & $  9.1_{-  0.9}^{+  0.8}$ & $  1.7_{-  0.5}^{+  0.5}$ \\
   3C 28             &   2.2 &   7.2 & $0.053_{-0.009}^{+0.015}$ & $ 1.29_{-0.19}^{+0.12}$ & $  9.1_{-  1.7}^{+  2.0}$ & $  2.8_{-  1.5}^{+  1.9}$ \\
   A133              &   3.0 &   3.3 & $0.048_{-0.005}^{+0.004}$ & $ 1.77_{-0.06}^{+0.09}$ & $ 13.5_{-  1.1}^{+  1.1}$ & $  2.2_{-  0.6}^{+  0.9}$ \\
   A223              &   2.0 &   6.7 & $0.019_{-0.003}^{+0.003}$ & $ 5.50_{-0.61}^{+0.80}$ & $ 76.3_{-  11.4}^{+ 13.4}$ & $ 24.0_{-  5.2}^{+  5.5}$ \\
   A262              &   2.0 &   0.7 & $0.067_{-0.019}^{+0.019}$ & $ 0.86_{-0.01}^{+0.01}$ & $  5.2_{-  1.0}^{+  1.0}$ & $  0.8_{-  0.2}^{+  0.2}$ \\
   A383              &   1.5 &   4.6 & $0.116_{-0.012}^{+0.010}$ & $ 2.03_{-0.26}^{+0.25}$ & $  8.5_{-  1.2}^{+  1.2}$ & $  1.8_{-  0.6}^{+  0.7}$ \\
   AWM 7             &   2.2 &   0.8 & $0.140_{-0.013}^{+0.015}$ & $ 1.18_{-0.07}^{+0.07}$ & $  4.4_{-  0.4}^{+  0.4}$ & $  0.9_{-  0.3}^{+  0.3}$ \\
   Perseus           &  11.8 &   4.2 & $0.150_{-0.005}^{+0.005}$ & $ 4.39_{-0.39}^{+0.46}$ & $ 15.5_{-  1.4}^{+  1.7}$ & $  2.5_{-  0.4}^{+  0.4}$ \\
   2A 0335+096       &   3.4 &   2.4 & $0.100_{-0.011}^{+0.011}$ & $ 1.32_{-0.09}^{+0.07}$ & $  6.1_{-  0.6}^{+  0.6}$ & $  1.4_{-  0.5}^{+  0.6}$ \\
   A478              &   1.5 &   2.6 & $0.197_{-0.015}^{+0.014}$ & $ 2.68_{-0.28}^{+0.28}$ & $  7.9_{-  0.9}^{+  0.9}$ & $  1.3_{-  0.3}^{+  0.4}$ \\
   A496              &   1.0 &   0.6 & $0.198_{-0.025}^{+0.031}$ & $ 1.20_{-0.15}^{+0.13}$ & $  3.5_{-  0.5}^{+  0.5}$ & $  0.7_{-  0.3}^{+  0.3}$ \\
   A520              &   7.7 &  25.2 & $0.005_{-0.001}^{+0.001}$ & $ 7.07_{-0.90}^{+3.24}$ & $253.8_{- 43.2}^{+120.4}$ & $118.0_{- 48.0}^{+ 66.5}$ \\
   MS 0735.6+7421    &   2.9 &  10.1 & $0.067_{-0.002}^{+0.002}$ & $ 3.18_{-0.24}^{+0.22}$ & $ 19.2_{-  1.5}^{+  1.4}$ & $  4.7_{-  0.6}^{+  0.7}$ \\
   PKS 0745-191      &   2.5 &   4.8 & $0.143_{-0.010}^{+0.010}$ & $ 2.62_{-0.36}^{+0.37}$ & $  9.5_{-  1.4}^{+  1.4}$ & $  2.1_{-  0.5}^{+  0.6}$ \\
   Hydra A           &   2.2 &   2.4 & $0.149_{-0.019}^{+0.014}$ & $ 2.59_{-0.48}^{+0.75}$ & $  9.2_{-  1.9}^{+  2.7}$ & $  2.0_{-  0.6}^{+  1.0}$ \\
   Zw 3146         &   1.6  &  6.8  & $0.177_{-0.007}^{+0.007}$ & $ 3.09_{-0.25}^{+0.27}$ & $  9.8_{-  0.8}^{+  0.9}$ & $  1.9_{-  0.3}^{+  0.3}$ \\
   A1068             &   2.0 &   4.9 & $0.149_{-0.011}^{+0.006}$ & $ 2.16_{-0.09}^{+0.42}$ & $  7.7_{-  0.5}^{+  1.5}$ & $  1.4_{-  0.3}^{+  0.5}$ \\
   A1361             &   2.0 &   4.2 & $0.044_{-0.010}^{+0.010}$ & $ 2.59_{-0.22}^{+0.16}$ & $ 20.7_{-  3.6}^{+  3.3}$ & $  5.7_{-  1.7}^{+  1.7}$ \\
   A1413             &   3.8 &   9.5 & $0.039_{-0.001}^{+0.002}$ & $ 6.14_{-1.00}^{+1.35}$ & $ 53.4_{-  8.8}^{+ 11.8}$ & $ 13.1_{-  2.6}^{+  2.9}$ \\
   M87               &   5.9 &   0.5 & $0.191_{-0.009}^{+0.009}$ & $ 0.94_{-0.02}^{+0.02}$ & $  2.8_{-  0.1}^{+  0.1}$ & $  0.4_{-  0.1}^{+  0.1}$ \\
   HCG 62            &   2.0 &   0.6 & $0.072_{-0.014}^{+0.014}$ & $ 0.67_{-0.01}^{+0.01}$ & $  3.9_{-  0.6}^{+  0.6}$ & $  0.6_{-  0.2}^{+  0.3}$ \\
   A1650             &   3.0 &   4.7 & $0.043_{-0.002}^{+0.002}$ & $ 4.26_{-0.90}^{+1.59}$ & $ 34.4_{-  7.4}^{+ 12.9}$ & $  8.1_{-  1.8}^{+  2.8}$ \\
   Coma              &   9.8 &   4.6 & $0.008_{-0.003}^{+0.003}$ & $ 2.75_{-1.61}^{+3.45}$ & $ 66.9_{- 42.6}^{+ 85.4}$ & $ 32.0_{- 31.3}^{+ 73.8}$ \\
   A1795             &   2.0 &   2.4 & $0.069_{-0.016}^{+0.016}$ & $ 2.74_{-0.37}^{+0.57}$ & $ 16.3_{-  3.4}^{+  4.3}$ & $  3.7_{-  1.4}^{+  1.7}$ \\
   A1835             &   3.2 &  12.7 & $0.110_{-0.003}^{+0.003}$ & $ 4.03_{-0.27}^{+0.27}$ & $ 17.6_{-  1.2}^{+  1.2}$ & $  3.0_{-  0.4}^{+  0.4}$ \\
   A1991             &   2.2 &   2.5 & $0.077_{-0.011}^{+0.011}$ & $ 0.77_{-0.02}^{+0.02}$ & $  4.2_{-  0.4}^{+  0.4}$ & $  0.6_{-  0.2}^{+  0.3}$ \\
   MS 1455.0+2232    &   1.3 &   5.3 & $0.095_{-0.020}^{+0.016}$ & $ 4.40_{-0.90}^{+1.36}$ & $ 21.1_{-  5.2}^{+  7.0}$ & $  2.8_{-  1.3}^{+  2.6}$ \\
   RXC J1504.1-0248  &   1.1 &   3.8 & $0.180_{-0.012}^{+0.011}$ & $ 6.87_{-2.53}^{+3.82}$ & $ 21.5_{-  8.0}^{+ 12.0}$ & $  3.2_{-  1.2}^{+  1.1}$ \\
   A2029             &   0.6 &   0.9 & $0.373_{-0.031}^{+0.036}$ & $ 2.88_{-0.22}^{+0.33}$ & $  5.6_{-  0.5}^{+  0.7}$ & $  0.5_{-  0.1}^{+  0.2}$ \\
   A2052             &   3.9 &   2.7 & $0.017_{-0.002}^{+0.002}$ & $ 0.71_{-0.08}^{+0.04}$ & $ 10.5_{-  1.4}^{+  1.0}$ & $  3.5_{-  1.8}^{+  7.8}$ \\
   A2065             &   2.0 &   2.7 & $0.037_{-0.011}^{+0.011}$ & $ 1.97_{-0.21}^{+0.35}$ & $ 17.8_{-  4.2}^{+  4.8}$ & $  6.6_{-  2.9}^{+  3.3}$ \\
   RX J1532.8+3021   &   2.0 &   9.6 & $0.107_{-0.009}^{+0.009}$ & $ 3.32_{-0.28}^{+0.29}$ & $ 14.8_{-  1.5}^{+  1.6}$ & $  3.3_{-  0.5}^{+  0.6}$ \\
   A2218             &   3.9 &  11.7 & $0.008_{-0.003}^{+0.003}$ & $ 3.44_{-0.56}^{+0.96}$ & $ 83.3_{- 25.9}^{+ 32.1}$ & $ 34.5_{- 18.1}^{+ 22.2}$ \\
   Hercules A        &   2.0 &   5.3 & $0.078_{-0.010}^{+0.010}$ & $ 2.04_{-0.21}^{+0.19}$ & $ 11.2_{-  1.6}^{+  1.5}$ & $  2.9_{-  0.8}^{+  0.8}$ \\
   A2244             &   2.0 &   3.5 & $0.046_{-0.008}^{+0.008}$ & $ 4.58_{-0.60}^{+0.92}$ & $ 35.6_{-  6.4}^{+  8.3}$ & $  6.6_{-  2.1}^{+  3.1}$ \\
   NGC 6338          &   1.5 &   0.8 & $0.236_{-0.018}^{+0.022}$ & $ 0.99_{-0.06}^{+0.05}$ & $  2.6_{-  0.2}^{+  0.2}$ & $  0.5_{-  0.1}^{+  0.1}$ \\
   RX J1720.2+2637   &   1.7 &   4.8 & $0.099_{-0.019}^{+0.017}$ & $ 2.97_{-0.61}^{+0.83}$ & $ 13.8_{-  3.3}^{+  4.2}$ & $  2.7_{-  1.5}^{+  2.1}$ \\
   MACS J1720.2+3536 &   2.2 &  11.7 & $0.076_{-0.006}^{+0.005}$ & $ 3.66_{-0.25}^{+0.34}$ & $ 20.4_{-  1.8}^{+  2.1}$ & $  3.5_{-  0.8}^{+  1.0}$ \\
   A2261             &   2.0 &   7.1 & $0.043_{-0.010}^{+0.010}$ & $ 5.76_{-0.92}^{+1.35}$ & $ 46.6_{-  10.9}^{+ 13.4}$ & $  7.3_{-  3.0}^{+  4.3}$ \\
   A2319             &   4.4 &   4.8 & $0.028_{-0.014}^{+0.014}$ & $11.32_{-3.41}^{+7.47}$ & $122.2_{- 55.0}^{+ 90.4}$ & $ 24.4_{- 14.3}^{+ 17.6}$ \\
   A2390             &   0.6 &   2.3 & $0.199_{-0.015}^{+0.010}$ & $ 3.39_{-0.86}^{+1.35}$ & $ 10.0_{-  2.6}^{+  4.0}$ & $  2.0_{-  0.6}^{+  0.6}$ \\
   A2409             &   3.0 &   7.6 & $0.013_{-0.008}^{+0.008}$ & $ 4.82_{-0.68}^{+0.98}$ & $ 88.5_{- 40.6}^{+ 42.6}$ & $ 21.4_{- 16.0}^{+ 18.6}$ \\
   A2597             &   3.4 &   5.5 & $0.075_{-0.005}^{+0.005}$ & $ 1.66_{-0.14}^{+0.12}$ & $  9.3_{-  0.9}^{+  0.8}$ & $  2.6_{-  0.6}^{+  0.6}$ \\
   A2626             &   2.0 &   2.1 & $0.063_{-0.014}^{+0.014}$ & $ 2.54_{-0.64}^{+0.75}$ & $ 16.0_{-  4.9}^{+  5.4}$ & $  4.1_{-  2.2}^{+  2.4}$ \\
   A2657             &   3.0 &   2.3 & $0.017_{-0.009}^{+0.009}$ & $ 3.84_{-0.87}^{+1.39}$ & $ 58.2_{- 24.6}^{+ 29.7}$ & $ 18.4_{- 14.1}^{+ 18.0}$ \\
   A2670             &   3.0 &   4.3 & $0.026_{-0.012}^{+0.012}$ & $ 3.38_{-0.65}^{+0.90}$ & $ 38.3_{- 15.9}^{+ 17.0}$ & $  6.6_{-  6.0}^{+  8.5}$ \\
\enddata
\tablenotetext{a}{Mean radius of the inner-most region; when elliptical regions were used, the equivalent redius ($r=[ab]^{1/2}$, where $a$ and $b$ are the semi-axes) is given.}
\end{deluxetable*}

The situation can be improved somewhat using the fact that the emissivity and hence surface brightness for a fully ionized plasma is a strong function of the density ($\epsilon \sim n_e^2$) to extrapolate the density closer to the core than is generally possible using spectral deprojection.  We used this method for 15 of the 46 objects with optical data; for the remaining objects, this method did not result in a significant decrease in the central region over that used in spectral deprojection (due either to reaching the resolution limit of $r \approx 1$ arcsec or to substructure in the cluster emission that resulted in unphysical emissivities when the central region became too small). 

We used the onion-peel deprojection method \citep[e.g.,][]{ameg07} to transform from X-ray surface brightness to density. We extracted the surface brightness profile between 0.5--7 keV in circular annuli with $\approx 200$--500 counts each, after background subtraction. When deprojecting the surface brightness profile, we neglected any emission from the cluster beyond the outermost radius of the surface brightness profile, which can result in errors in the outer parts of the density profile if there is still significant cluster emission beyond this radius. However, as we are interested only in the innermost portions of the profiles, this effect should not affect the central density that we derive. 

We then normalized the resulting density profile to the coarser density profile derived using the PROJCT model, and we used this normalized density profile to estimate the central density. Errors on the derived central densities were estimated using a Monte Carlo  technique, wherein the deprojection was repeated 100 times, with the surface brightness in each annulus drawn randomly from a poisson distribution given by the total counts in each annulus (before background subtraction). We also included errors in the central density due to uncertainties in the normalization by assuming, conservatively, that this error is equal to the error of the inner density derived from spectral deprojection. Finally, when calculating the central cooling time and entropy using this density, we assumed that the temperature and abundance of the gas are constant in the inner region used in spectral deprojection and equal to the derived emission-weighted values.   Table \ref{T:ICM_central} lists the resulting central properties (and 1-$\sigma$ errors), derived as close as possible to the core (after the exclusion of any non-thermal point sources). 

Lastly, to make comparisons of the ICM properties between objects at a single physical radius, we used linear interpolation of the profiles in $\log - \log$ space, using the mean radius of each annulus. We chose a radius of 12 kpc, as this radius was the smallest physical radius that could be achieved across most of our sample (excluding only A520; the mean radius of A1835 lies at 12.7 kpc, but the properties at this radius should differ only slightly from those at 12 kpc).  We also interpolated the cooling-time profiles in the same way to find the cooling time at the radius corresponding to the distance between the X-ray core and the CDG's core (denoted $\Delta r$ and listed in Table \ref{T:sample}). Table \ref{T:ICM_12kpc} lists the interpolated ICM properties.

\begin{deluxetable*}{lccccccc}
\tablewidth{0pt}
\tablecaption{Interpolated ICM Properties. \label{T:ICM_12kpc}}
\tablehead{
	\colhead{} & \multicolumn{4}{c}{$r=12$ kpc} & \colhead{} & \colhead{$r=\Delta r$\tablenotemark{a}} \\
	\cline{2-5} \cline{7-7}
	\colhead{} & \colhead{$n_e$} & \colhead{$kT$} & \colhead{Entropy} & \colhead{$t_{\rm{cool}}$} & \colhead{}  & \colhead{$t_{\rm{cool}}$}\\
	\colhead{System} & \colhead{(cm$^{-3}$)} & \colhead{(keV)} & \colhead{(keV cm$^2$)} & \colhead{(10$^8$ yr)} & \colhead{}  & \colhead{(10$^8$ yr)} 
 }
\startdata
   A85               & $0.039_{-0.001}^{+0.001}$ & $ 2.87_{-0.09}^{+0.11}$ & $ 25.1_{-  0.9}^{+  1.1}$ & $  6.4_{-  0.5}^{+  0.6}$ &  & $  3.1_{-  0.4}^{+  0.4}$ \\
   3C 28             & $0.039_{-0.007}^{+0.013}$ & $ 1.65_{-0.16}^{+0.11}$ & $ 14.3_{-  1.4}^{+  1.7}$ & $  4.8_{-  1.3}^{+  1.6}$ &  & $  5.2_{-  1.2}^{+  1.6}$ \\
   A133              & $0.024_{-0.003}^{+0.002}$ & $ 2.19_{-0.04}^{+0.05}$ & $ 26.0_{-  0.7}^{+  0.7}$ & $  6.5_{-  0.5}^{+  0.6}$ &  & $  3.8_{-  0.5}^{+  0.8}$ \\
   A223              & $0.013_{-0.003}^{+0.003}$ & $ 5.50_{-0.61}^{+0.80}$ & $100.4_{- 10.7}^{+ 12.6}$ & $ 36.6_{- 5.1}^{+ 5.3}$ &  & $120.0_{- 19.5}^{+ 21.1}$ \\
   A262              & $0.012_{-0.001}^{+0.001}$ & $ 1.53_{-0.01}^{+0.01}$ & $ 28.5_{-  0.4}^{+  0.4}$ & $  8.8_{-  0.5}^{+  0.5}$ &  & $  1.1_{-  0.3}^{+  0.3}$ \\
   A383              & $0.054_{-0.008}^{+0.007}$ & $ 2.54_{-0.19}^{+0.18}$ & $ 17.7_{-  1.1}^{+  1.0}$ & $  4.4_{-  0.6}^{+  0.6}$ &  & $  1.8_{-  0.6}^{+  0.7}$ \\
   AWM 7             & $0.014_{-0.001}^{+0.001}$ & $ 2.93_{-0.10}^{+0.14}$ & $ 49.4_{-  2.0}^{+  2.8}$ & $ 16.1_{-  1.7}^{+  2.0}$ &  & $  0.9_{-  0.3}^{+  0.3}$ \\
   Perseus           & $0.061_{-0.002}^{+0.002}$ & $ 3.41_{-0.19}^{+0.23}$ & $ 22.0_{-  0.7}^{+  0.8}$ & $  5.5_{-  0.2}^{+  0.2}$ &  & $  2.5_{-  0.4}^{+  0.4}$ \\
   2A 0335+096       & $0.055_{-0.003}^{+0.003}$ & $ 1.57_{-0.03}^{+0.02}$ & $ 10.8_{-  0.2}^{+  0.2}$ & $  3.1_{-  0.2}^{+  0.2}$ &  & $  2.9_{-  0.2}^{+  0.2}$ \\
   A478              & $0.072_{-0.009}^{+0.009}$ & $ 3.61_{-0.17}^{+0.18}$ & $ 20.9_{-  0.6}^{+  0.6}$ & $  4.6_{-  0.2}^{+  0.2}$ &  & $  1.3_{-  0.3}^{+  0.4}$ \\
   A496              & $0.030_{-0.001}^{+0.001}$ & $ 2.37_{-0.02}^{+0.02}$ & $ 24.7_{-  0.2}^{+  0.2}$ & $  6.9_{-  0.2}^{+  0.2}$ &  & $  0.7_{-  0.3}^{+  0.3}$ \\
   A520              & \nodata & \nodata & \nodata & \nodata &  & $224.8_{- 17.5}^{+ 17.8}$ \\
   MS 0735.6+7421    & $0.057_{-0.002}^{+0.002}$ & $ 3.24_{-0.23}^{+0.21}$ & $ 21.9_{-  1.5}^{+  1.3}$ & $  5.6_{-  0.6}^{+  0.6}$ &  & $  4.7_{-  0.6}^{+  0.7}$ \\
   PKS 0745-191      & $0.094_{-0.006}^{+0.006}$ & $ 3.15_{-0.21}^{+0.22}$ & $ 15.3_{-  0.8}^{+  0.9}$ & $  3.5_{-  0.3}^{+  0.3}$ &  & \nodata \\
   Hydra A           & $0.046_{-0.004}^{+0.003}$ & $ 2.83_{-0.10}^{+0.16}$ & $ 22.1_{-  0.5}^{+  0.6}$ & $  6.2_{-  0.2}^{+  0.3}$ &  & \nodata \\
   Zw 3146           & $0.112_{-0.006}^{+0.005}$ & $ 3.34_{-0.20}^{+0.21}$ & $ 14.4_{-  0.7}^{+  0.8}$ & $  3.0_{-  0.2}^{+  0.3}$ &  & \nodata \\
   A1068             & $0.069_{-0.006}^{+0.003}$ & $ 2.58_{-0.07}^{+0.24}$ & $ 15.4_{-  0.6}^{+  1.0}$ & $  3.5_{-  0.3}^{+  0.4}$ &  &\nodata \\
   A1361             & $0.032_{-0.005}^{+0.005}$ & $ 2.59_{-0.22}^{+0.16}$ & $ 26.0_{-  2.3}^{+  2.0}$ & $  8.0_{-  1.3}^{+  1.3}$ &  & $  5.8_{-  1.8}^{+  1.7}$ \\
   A1413             & $0.034_{-0.001}^{+0.002}$ & $ 6.19_{-0.93}^{+1.25}$ & $ 59.0_{-  8.2}^{+ 11.0}$ & $ 15.0_{-  2.4}^{+  2.7}$ &  & $ 13.1_{-  2.6}^{+  2.9}$ \\
   M87               & $0.019_{-0.001}^{+0.001}$ & $ 1.73_{-0.02}^{+0.02}$ & $ 24.2_{-  0.3}^{+  0.3}$ & $  8.6_{-  0.4}^{+  0.4}$ &  & $  0.4_{-  0.1}^{+  0.1}$ \\
   HCG 62            & $0.007_{-0.001}^{+0.001}$ & $ 0.82_{-0.01}^{+0.01}$ & $ 23.1_{-  0.9}^{+  0.9}$ & $  6.3_{-  1.0}^{+  1.1}$ &  & $  0.8_{-  0.2}^{+  0.3}$ \\
   A1650             & $0.026_{-0.001}^{+0.001}$ & $ 5.27_{-0.62}^{+1.06}$ & $ 60.0_{-  5.6}^{+  9.0}$ & $ 15.5_{-  1.6}^{+  2.1}$ &  & $  8.1_{-  1.8}^{+  2.8}$ \\
   Coma              & $0.006_{-0.002}^{+0.002}$ & $ 5.05_{-1.40}^{+2.56}$ & $154.0_{- 43.2}^{+ 68.2}$ & $ 71.8_{- 25.5}^{+ 52.7}$ &  & \nodata \tablenotemark{b} \\
   A1795             & $0.042_{-0.003}^{+0.003}$ & $ 3.25_{-0.22}^{+0.34}$ & $ 26.9_{-  1.5}^{+  2.1}$ & $  7.2_{-  0.8}^{+  0.9}$ &  & $  5.4_{-  2.0}^{+  2.3}$ \\
   A1835             & $0.110_{-0.003}^{+0.003}$ & $ 4.03_{-0.27}^{+0.27}$ & $ 17.6_{-  1.2}^{+  1.2}$ & $  3.0_{-  0.4}^{+  0.4}$ &  & \nodata \\
   A1991             & $0.028_{-0.001}^{+0.001}$ & $ 1.42_{-0.02}^{+0.02}$ & $ 15.3_{-  0.3}^{+  0.3}$ & $  4.1_{-  0.4}^{+  0.4}$ &  & $  2.7_{-  0.3}^{+  0.3}$ \\
   MS 1455.0+2232    & $0.077_{-0.012}^{+0.009}$ & $ 3.82_{-0.52}^{+0.79}$ & $ 21.1_{-  3.0}^{+  4.0}$ & $  3.9_{-  0.8}^{+  1.5}$ &  & $  3.0_{-  1.2}^{+  2.4}$ \\
   RXC J1504.1-0248  & $0.150_{-0.005}^{+0.004}$ & $ 4.70_{-1.03}^{+1.55}$ & $ 16.7_{-  3.2}^{+  4.9}$ & $  2.9_{-  0.5}^{+  0.5}$ &  & $  3.1_{-  1.0}^{+  1.0}$ \\
   A2029             & $0.057_{-0.011}^{+0.013}$ & $ 4.83_{-0.10}^{+0.13}$ & $ 32.6_{-  0.5}^{+  0.6}$ & $  6.0_{-  0.2}^{+  0.2}$ &  & \nodata \\
   A2052             & $0.024_{-0.001}^{+0.001}$ & $ 1.53_{-0.02}^{+0.02}$ & $ 18.5_{-  0.4}^{+  0.3}$ & $  6.1_{-  0.5}^{+  2.1}$ &  & \nodata \\
   A2065             & $0.019_{-0.003}^{+0.002}$ & $ 2.63_{-0.18}^{+0.30}$ & $ 36.3_{-  2.8}^{+  3.9}$ & $ 15.0_{-  2.4}^{+  3.0}$ &  & $ 15.0_{-  6.2}^{+  7.2}$ \\
   RX J1532.8+3021   & $0.094_{-0.007}^{+0.007}$ & $ 3.32_{-0.28}^{+0.29}$ & $ 16.1_{-  1.3}^{+  1.3}$ & $  3.7_{-  0.5}^{+  0.5}$ &  & $  5.3_{-  0.8}^{+  0.8}$ \\
   A2218             & $0.008_{-0.005}^{+0.005}$ & $ 3.44_{-0.56}^{+0.96}$ & $ 83.4_{- 25.6}^{+ 31.7}$ & $ 34.6_{- 18.0}^{+ 22.1}$ &  & $ 38.3_{- 17.7}^{+ 22.8}$ \\
   Hercules A        & $0.036_{-0.002}^{+0.002}$ & $ 2.05_{-0.21}^{+0.19}$ & $ 18.7_{-  2.1}^{+  1.9}$ & $  6.2_{-  1.6}^{+  1.5}$ &  & $  6.1_{-  1.6}^{+  1.5}$ \\
   A2244             & $0.022_{-0.002}^{+0.002}$ & $ 5.25_{-0.46}^{+0.70}$ & $ 66.5_{-  5.3}^{+  7.4}$ & $ 17.4_{-  2.1}^{+  3.3}$ &  & $  9.7_{-  2.7}^{+  4.2}$ \\
   NGC 6338          & $0.015_{-0.001}^{+0.001}$ & $ 1.95_{-0.08}^{+0.03}$ & $ 31.9_{-  1.6}^{+  1.2}$ & $  9.0_{-  1.6}^{+  1.7}$ &  & $  0.5_{-  0.1}^{+  0.1}$ \\
   RX J1720.2+2637   & $0.061_{-0.013}^{+0.011}$ & $ 3.61_{-0.42}^{+0.57}$ & $ 23.4_{-  2.3}^{+  2.9}$ & $  5.3_{-  1.0}^{+  1.5}$ &  & $  4.1_{-  1.3}^{+  1.8}$ \\
   MACS J1720.2+3536 & $0.074_{-0.006}^{+0.005}$ & $ 3.69_{-0.25}^{+0.34}$ & $ 20.9_{-  1.7}^{+  2.1}$ & $  3.6_{-  0.8}^{+  1.0}$ &  & $  3.5_{-  0.8}^{+  1.0}$ \\
   A2261             & $0.034_{-0.004}^{+0.003}$ & $ 5.76_{-0.92}^{+1.35}$ & $ 55.2_{- 9.2}^{+ 12.6}$ & $  9.3_{-  3.1}^{+  4.8}$ &  & $  9.5_{-  3.3}^{+  5.2}$ \\
   A2319             & $0.024_{-0.009}^{+0.009}$ & $11.32_{-3.41}^{+7.47}$ & $134.3_{- 39.1}^{+67.9}$ & $ 28.1_{- 10.1}^{+ 12.9}$ &  & $ 45.8_{- 18.9}^{+ 26.9}$ \\
   A2390             & $0.064_{-0.009}^{+0.006}$ & $ 4.38_{-0.50}^{+0.77}$ & $ 27.5_{-  1.7}^{+  2.4}$ & $  6.4_{-  0.4}^{+  0.4}$ &  & $  2.8_{-  0.6}^{+  0.6}$ \\
   A2409             & $0.011_{-0.006}^{+0.006}$ & $ 4.82_{-0.68}^{+0.98}$ & $94.8_{- 31.3}^{+33.0}$ & $ 23.7_{- 12.6}^{+ 14.8}$ &  & $ 28.2_{- 14.0}^{+ 18.9}$ \\
   A2597             & $0.048_{-0.003}^{+0.003}$ & $ 2.37_{-0.11}^{+0.10}$ & $ 17.9_{-  0.9}^{+  0.8}$ & $  4.8_{-  0.5}^{+  0.5}$ &  & $  2.6_{-  0.6}^{+  0.6}$ \\
   A2626             & $0.019_{-0.002}^{+0.002}$ & $ 2.59_{-0.28}^{+0.33}$ & $ 35.9_{-  2.7}^{+  3.0}$ & $ 12.7_{-  1.7}^{+  1.8}$ &  & $  4.9_{-  2.6}^{+  2.8}$ \\
   A2657             & $0.009_{-0.001}^{+0.002}$ & $ 4.68_{-0.83}^{+1.37}$ & $112.2_{- 22.4}^{+ 36.1}$ & $ 44.4_{- 15.9}^{+ 21.6}$ &  & $ 21.6_{- 16.4}^{+ 21.1}$ \\
   A2670             & $0.008_{-0.003}^{+0.002}$ & $ 3.53_{-0.61}^{+0.84}$ & $ 85.6_{- 20.5}^{+ 23.7}$ & $ 23.4_{- 14.1}^{+ 21.8}$ &  & $ 19.3_{- 15.1}^{+ 23.4}$ \\
\enddata
\tablenotetext{a}{The $\Delta r$ radius is the projected distance between the X-ray core and the CDG core (see Table~\ref{T:sample}).}
\tablenotetext{b}{The Coma data lack sufficient counts to derive a cooling time at the CDG's radius.}
\end{deluxetable*}

\subsection{Analysis of Optical Data}
We identified the CDG as the most extended galaxy within the field, which generally lies near the X-ray core. However, when the cluster contains more than one very large galaxy and it is unclear which is the dominant galaxy, we chose as the CDG the galaxy with the largest integrated $K$-band magnitude from the 2MASS catalog. Below we note any unusual characteristics of the CDG or image.

\begin{description}
 \item[A85:] The CDG has extended optical line emission \citep{hu85,fish95}.
 \item[3C 28:]\vspace{-2.5mm} An excess of UV flux has been detected in the CDG by \citet{will02} and was ascribed by them to active star formation.
 \item[A262:]\vspace{-2.5mm} The CDG has a large, central dust lane and a bright star nearby which were masked.
 \item[Perseus:]\vspace{-2.5mm} The Perseus CDG (NGC 1275) is well known for its blue, emission-line filaments, dust lanes, and the foreground high-velocity system, all of which were masked before analysis. Significant star formation was detected by \citet{mcna89,roma87,smit92}.
 \item[2A 0335+096:]\vspace{-2.5mm} There is a nearby bright star in the image that results in large areas of the galaxy being masked. Significant star formation was detected by \citet{roma88}.
 \item[A478:]\vspace{-2.5mm} \citet{card98} found spectral evidence for significant star formation in the CDG.
 \item[A520:]\vspace{-2.5mm} The cluster appears to have no well-defined optical core \citep{dahl02}, and instead has several concentrations of galaxies, none of which coincide with the X-ray core. \citet{craw99} note that there are three dominant galaxies and chose the SW one as the CDG, as did we.
 \item[M87:]\vspace{-2.5mm} The optical jet and central region, associated with the AGN, were masked before analysis.
 \item[HCG 62:]\vspace{-2.5mm} The dominant galaxy of this compact group has a large, nearby companion galaxy which was masked before analysis.
 \item[A1795:]\vspace{-2.5mm} The CDG is well known to harbor a tail of blue emission that is thought to be due to star formation triggered by compression of the gas by the radio source \citep{mcna96a,mcna96b,pink96,odea04}. The $R$- and $I$-band images were affected by scattered light, which was masked as thoroughly as possible.
 \item[A1991:]\vspace{-2.5mm} As with A1795, the long-wavelength images were affected by scattered light, requiring masking, and the seeing was poor. \citet{mcna89} report spectral evidence for a small SFR, and \citet{mcna92} report a positive $U-I$ color gradient in the core.
 \item[RXC J1504.1-0248:]\vspace{-2.5mm} The $U$-band image shows an elongated blue region in the core, extending 6 arcsec in length.
 \item[A2065:]\vspace{-2.5mm} This cluster has two dominant galaxies. We have analyzed the southern galaxy, which appears to be associated with the X-ray core \citep{chat06}.
 \item[RX J1532.5+3021:]\vspace{-2.5mm} The CDG is known to be very blue \citep{dahl02}. The $U-I$ colormap shows a very blue region offset slightly to the south of the core and extending 4--8 arcsec in radius.
 \item[Hercules A:]\vspace{-2.5mm} The CDG has extended optical line emission \citep{tadh93}. The CDG core is the southeastern of the two surface-brightness peaks.
 \item[A2244:]\vspace{-2.5mm} A nearby bright star required large areas of masking.
 \item[RX J1720.2+2637:]\vspace{-2.5mm} The CDG has a blue central region some 13 arcsec in radius.
 \item[MACS J1720.2+3536:]\vspace{-2.5mm} The $U-I$ colormap shows blue emission extending to a radius of 2.5 arcsec.
 \item[A2390:]\vspace{-2.5mm} The $U$-band image shows a very blue, elongated region in the core, extending SE to NW some 4--5 arcsec in length. 
 \item[A2597:]\vspace{-2.5mm} The CDG possesses knots of star formation \citep{koek02,odea04}.
\end{description}
 
\subsubsection{Surface Brightness Profiles}\label{S:profiles}
The master frames, created by stacking the individual frames for each object (see Section \ref{S:stacking}), were first registered to a common coordinate grid using \textit{wregister} in IRAF. Next, the master frames were PSF-matched using the \textit{psfmatch} task to the lowest-resolution frame of the set (typically the $U$-band frame). This step is necessary before spatial comparisons between frames taken in different bands can be made. However, since the bands are fairly broad ($FWHM \sim 1500$ \AA) and the stars used for the matching may have different colors than the CDG, some effect due to PSF mismatches will remain in the data.  
\begin{figure*}
\begin{center}
$\begin{array}{l@{\hspace{0.5cm}}r}
\includegraphics[width=0.48\textwidth]{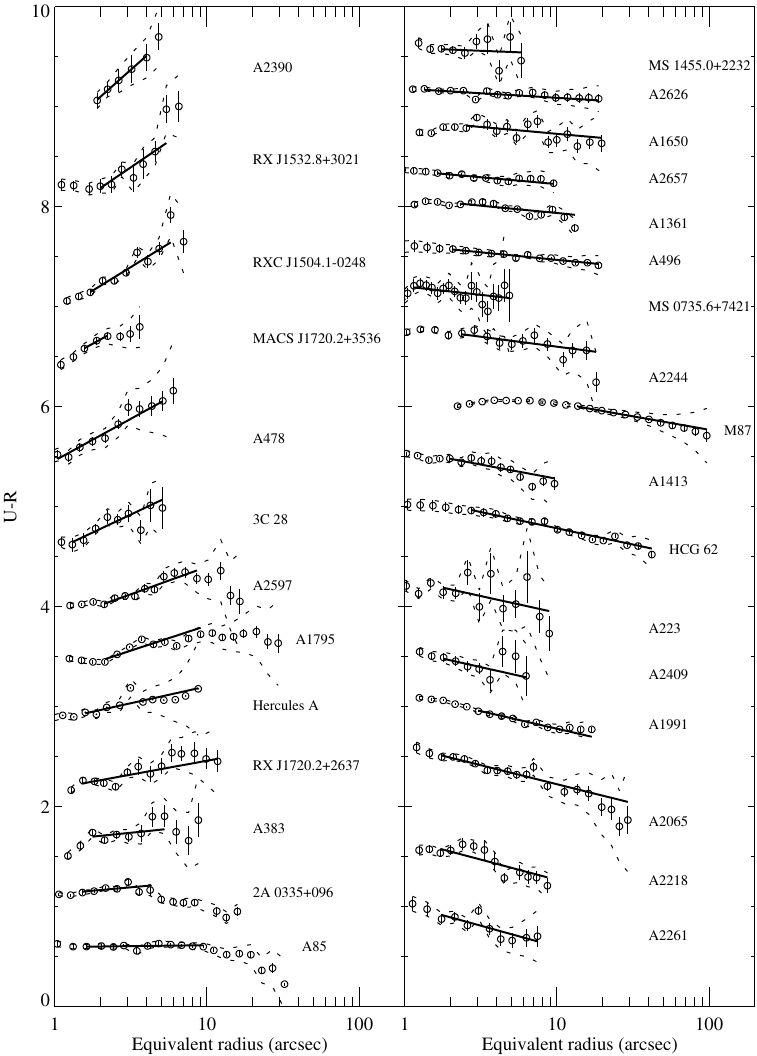} & \includegraphics[width=0.48\textwidth]{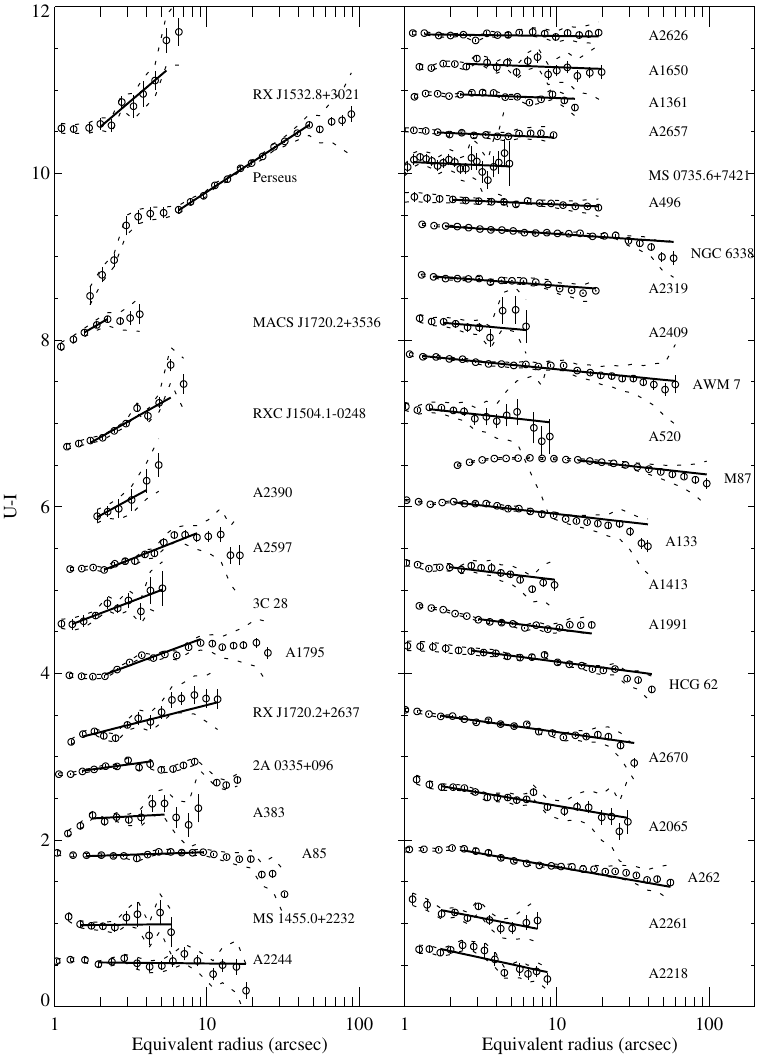}
\end{array}$
\end{center}
\caption{$U-R$ (\textit{left}) and $U-I$ (\textit{right}) color versus the equivalent radius ($r=[ab]^{1/2},$ where $a$ and $b$ are the semiaxes), ordered by decreasing gradient. The colors for each object have been shifted by arbitrary values. The error bars shown for each point are the statistical errors; the dashed lines show the total (statistical plus systematical) errors. The best-fit gradients (see Section \ref{S:Color_gradients}) are overplotted as solid lines between the inner and outer radii used in the fit. \label{F:UR_color_profiles}}
\end{figure*}

Surface brightness profiles were constructed from the PSF-matched, registered master frames using the \textit{ellipse} task in IRAF. This task works by fitting ellipses to isophotes of the surface brightness distribution of a galaxy and calculating the mean surface brightness along the ellipse \citep[see][]{jedr87}. Clipping routines are used to reject stars, superimposed galaxies, or other source of contamination. Additionally, such contaminating objects as were visible were masked out by eye before fitting. The ellipticities and position angles were allowed to vary from ellipse to ellipse to reflect changes in the underlying galaxy. Typically, the surface brightness distributions of CDGs become more elliptical at larger radii \citep[e.g.,][]{pate06}. The ellipse centroids were not allowed to vary, to make comparisons between profiles in different bands possible.  The centroids were determined from the long-wavelength images ($R$ or $I$), which are generally smooth and relaxed (reflecting the old stellar populations of the galaxy). Centroid positions were calculated using \textit{imexamine} in IRAF, which finds the centroid using Gaussian fits to the radial surface brightness profile. 

The ellipse centers, ellipticities, and position angles were fixed to those resulting from a fit to the $R$- or $I$-band image, whichever was of higher signal-to-noise, to make it possible to compare profiles in different bands. We note that fixing the $U$-band ellipse properties to those of a longer-wavelength band will tend to reduce the $U$-band surface brightness along the ellipse if the $U$-band emission has significantly different preferred ellipticities or position angles. An example of such a case is A2390, which has a bar of very blue emission near its center that is clearly different in shape from the $I$-band emission in the same region. The net effect of our procedure is to dilute somewhat the signatures of star formation in our profiles; however, we estimate that the effect is generally small. The resulting color profiles for the optical sample are shown in Figure~\ref{F:UR_color_profiles}. 

\subsubsection{Comparison with Other Studies and between Observing Runs}\label{S:Optical_comparison}

\begin{figure}
\plotone{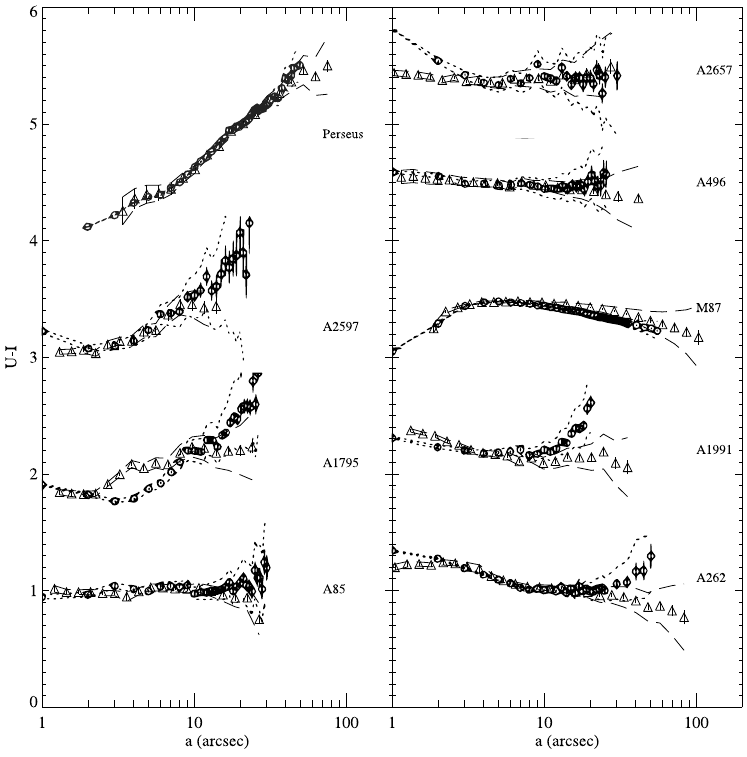}
\caption{Comparison of our $U-I$ color profiles (\textit{triangles}) to those of \citet{mcna92} (\textit{circles}) for the objects common to both samples. \label{F:MO92_overlays}}
\end{figure}
As a check for systematic errors in our reduction and analysis, we can compare our color profiles both to profiles from the literature and between different observing runs. In Figure \ref{F:MO92_overlays}, we compare our $U-I$ profiles with those from Figure 5 of \citet{mcna92}. In general, the profiles agree well within the errors over the inner 10-20 arcsec. At the extreme inner radii there is often some discrepancy, probably due to seeing effects at $a \lesssim 2$ arcsec, as the images of \citet{mcna92} were not matched between bands for different seeing. At the extreme outer radii ($a \gtrsim 10-20$ arcsec) our profiles do not match well, as those of  \citet{mcna92} often show a steep positive gradient that ours do not (particularly apparent in the profiles of A1991 and A262). This effect may be due to misestimation of the sky by \citet{mcna92}, as they used a much smaller CCD (FOV$\sim 2.5 \times 2.5$ arcmin compared to our smallest FOV of $9.4 \times 9.4$ arcmin) which may have resulted in sky boxes with some contamination from the central galaxy's halo. Otherwise, the only significant discrepancy is in the $3 < a < 6$ arcsec region of the profile of A1795. We believe the bluer colors that \citet{mcna92} find in this region are due to differences in the annuli used to measure the surface brightness, as the $U$-band image shows a linear blue feature in this region, the color of which would be diluted if fairly circular annuli are used (as we have done).

\begin{figure}
\plottwo{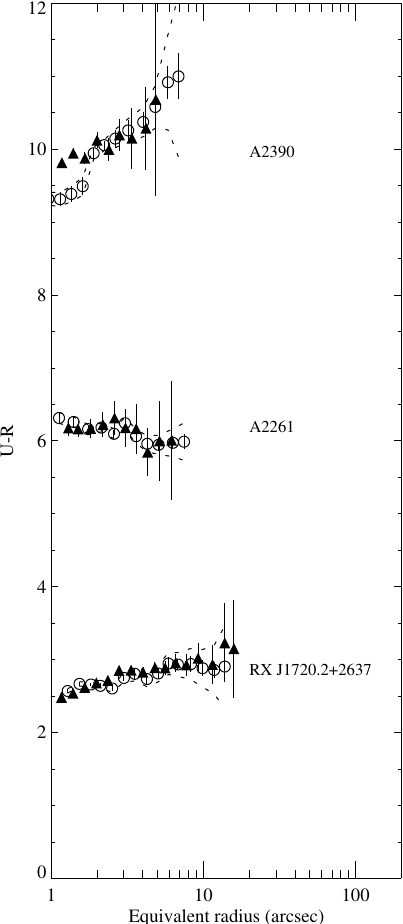}{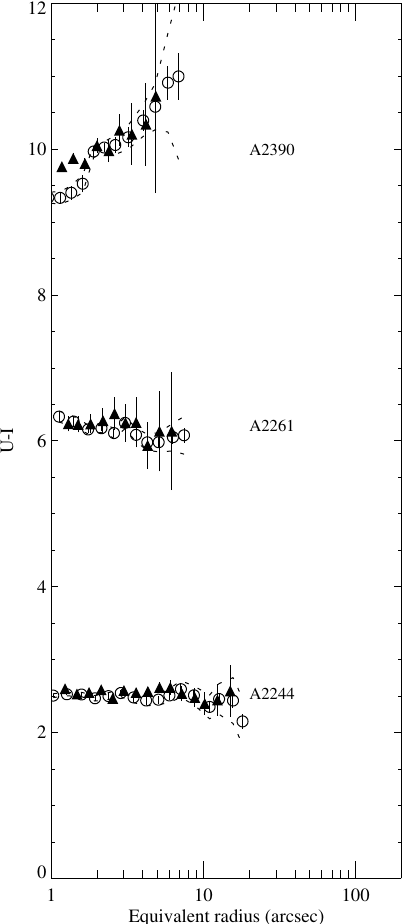}
\caption{Comparison of $U-R$ color profiles (\textit{left}) and $U-I$ color profiles (\textit{right}) from different observing runs: A2390 and RX J1720.2+2637 -- MDM (\textit{triangles}) and KPNO (\textit{circles}), A2261 and A2244 -- MDM September 2005 (\textit{triangles}) and MDM May 2006 (\textit{circles}). \label{F:internal_comparison}}
\end{figure}
As a further check of consistency, we also compare profiles taken from observations made during different runs and at different telescopes. In Figure \ref{F:internal_comparison} we show profiles for three objects that we observed twice. A2390 was observed with the MDM 2.4-m telescope and the KPNO 4-m telescope; A2261 and A2244 were observed twice with MDM. The profiles agree well, with the exception of the inner 2 arcsec of the A2390 profiles, which is probably due to seeing differences between the two observations. 

\subsection{Color Gradients and $\Delta$ Colors}\label{S:Color_gradients}
Color gradients were derived for the $U-I$ and $U-R$ color profiles using a least-squares fitting routine in the MPFIT package. The color gradient is defined as the change in color (in magnitudes) over the corresponding change in $\log(r)$ (in dex); e.g., for $U-I$:
\begin{equation}
 G(U-I) = \frac{d(U-I)}{d\log(r)}.
\end{equation}
The following function was fit to the data in color-$\log(r)$ space:
\begin{equation}
 U-I=G(U-I) \log(r) + b,
\end{equation}
where $G(U-I)$ is the color gradient and $b$ is the intercept. The data were weighted by their total (systematic plus statistical) error, and errors on the gradient were returned by the covariance matrix of the best fit. To eliminate possible effects from the PSF on the fitted gradient, the fits were restricted to radii greater than twice the radius of the FWHM of the PSF. 

The outer radius used in the fit was set differently for objects with blue cores and for those without. For blue objects, the outer radius was set to the approximate point at which the excess blue emission ends, so that the resulting gradient would trace the star-forming region only. The edge of the blue emission was determined by examining the $U-I$ or $U-R$ colormaps of the galaxy, made by dividing the $U$-band image by either the $R$- or $I$-band image. For red objects, the outer radius was set to the radius at which the total errors reach 0.5 mag to avoid regions where sky-subtraction errors significantly affect the profile (however, since the data were weighted by their errors, this choice has little effect on the measured gradients). See Figure \ref{F:UR_color_profiles} for plots of the color profiles with the best-fit gradients overlaid. Table~\ref{T:gradients} lists the color gradients and the radial range over which they were derived.
\begin{figure}
 \plotone{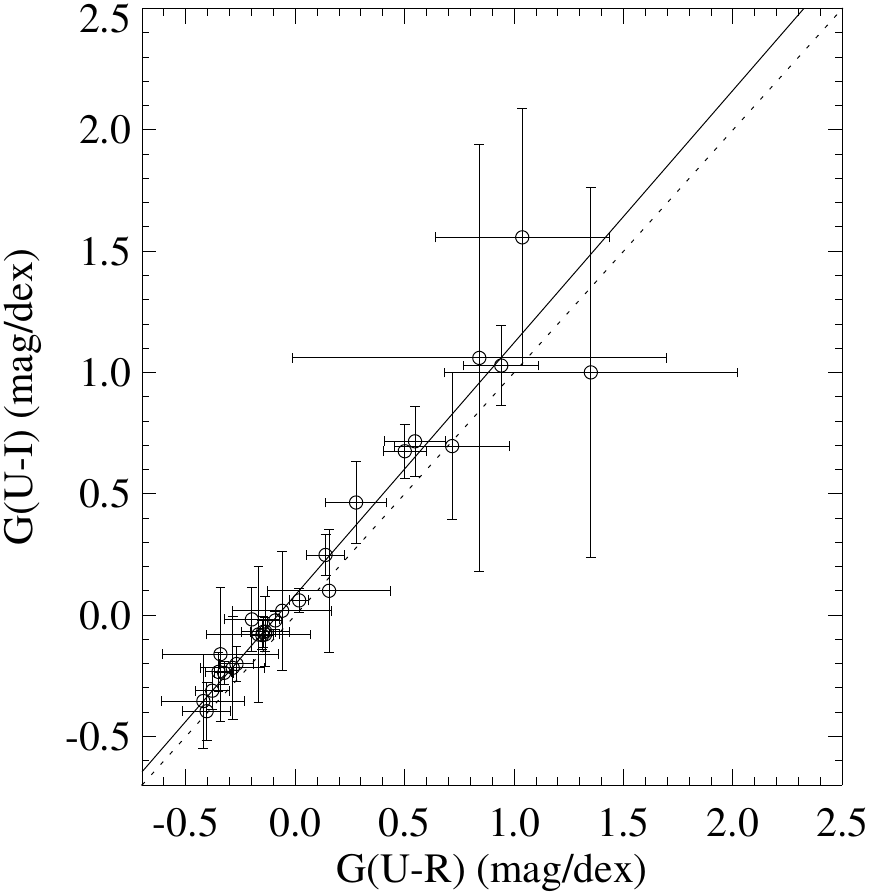}
\caption{The $U-I$ color gradient versus the $U-R$ color gradient. The dotted line shows equality between the two gradients, while the solid line shows the best-fit straight line. \label{F:GU_R_vs_GU_I}}
\end{figure}

\tabletypesize{\tiny}
\def\arraystretch{1.05}
\begin{deluxetable*}{lccccccc}
\tablewidth{0pt}
\tablecaption{CDG Color Gradients and $\Delta$ Colors. \label{T:gradients}}
\tablehead{
	\colhead{} &  \multicolumn{3}{c}{Color Gradients} & \colhead{} & \multicolumn{3}{c}{$\Delta$ Colors} \\
	\cline{2-4}  \cline{6-8}
	\colhead{} &  \colhead{$G(U-R)$} & \colhead{$G(U-I)$} & \colhead{Range} & \colhead{} & \colhead{$\Delta(U-R)$} & \colhead{$\Delta(U-I)$} & \colhead{Range} \\
	\colhead{System} &  \colhead{(mag dex$^{-1}$)} & \colhead{(mag dex$^{-1}$)} & \colhead{(arcsec)} & \colhead{} & \colhead{(mag)} & \colhead{(mag)} & \colhead{(kpc)} }
\startdata
   A85               & $ 0.02 \pm  0.04$ & $ 0.06 \pm  0.05$ &  1.6-9.5 &  & $-0.03 \pm  0.05$ & $ 0.00 \pm  0.05$ & 5-10 \\
                     & & & & & $-0.10 \pm  0.10$ & $-0.08 \pm  0.10$ & 5-20 \\
   3C 28             & $ 0.72 \pm  0.26$ & $ 0.70 \pm  0.30$ &  1.3-5.1 &  & $ 0.26 \pm  0.14$ & $ 0.24 \pm  0.15$ & 5-10 \\
   A133              &           \nodata & $-0.21 \pm  0.04$ &  1.8-39.6 &  &           \nodata & $-0.09 \pm  0.05$ & 5-10 \\
                       & & & & & \nodata & $-0.19 \pm  0.10$ & 5-20 \\
   A223              & $-0.34 \pm  0.31$ &           \nodata &  1.5-8.9 &  & $-0.14 \pm  0.18$ &           \nodata & 5-10 \\
                     & & & & & $-0.06 \pm  0.39$ & \nodata & 5-20 \\
   A262              &           \nodata & $-0.32 \pm  0.03$ &  2.0-176.4 &  &           \nodata & $-0.06 \pm  0.10$ & 5-10 \\
                       & & & & & \nodata & $-0.19 \pm  0.25$ & 5-20 \\
   A383              & $ 0.15 \pm  0.28$ & $ 0.10 \pm  0.25$ &  1.8-5.3 &  & \nodata & \nodata & \nodata  \\
   AWM 7             &           \nodata & $-0.18 \pm  0.03$ &  1.1-84.8 &  &           \nodata & $-0.09 \pm  0.18$ & 5-10 \\
                       & & & & & \nodata & $-0.17 \pm  0.45$ & 5-20 \\
   Perseus           &           \nodata & $ 1.19 \pm  0.07$ &  6.5-47.0 &  &           \nodata & $ 0.38 \pm  0.05$ & 5-10 \\
                       & & & & & \nodata & $ 0.59 \pm  0.17$ & 5-20 \\
   2A 0335+096       & $ 0.14 \pm  0.08$ & $ 0.25 \pm  0.09$ &  1.5-4.2 &  & $-0.12 \pm  0.08$ & $-0.22 \pm  0.08$ & 5-10 \\
   A478              & $ 0.84 \pm  0.25$ &           \nodata &  1.0-5.1 &  & $ 0.16 \pm  0.53$ &           \nodata & 5-10 \\
   A496              & $-0.15 \pm  0.05$ & $-0.08 \pm  0.06$ &  1.7-18.7 &  & $-0.05 \pm  0.08$ & $-0.04 \pm  0.09$ & 5-10 \\
                     & & & & & $-0.08 \pm  0.22$ & $-0.11 \pm  0.24$ & 5-20 \\
   A520              &           \nodata & $-0.20 \pm  0.28$ &  1.2-9.0 &  &           \nodata & $-0.13 \pm  0.16$ & 5-10 \\
                       & & & & & \nodata & $-0.13 \pm  0.48$ & 5-20 \\
   MS 0735.6+7421    & $-0.17 \pm  0.24$ & $-0.08 \pm  0.28$ &  1.0-4.9 &  & $-0.03 \pm  0.24$ & $ 0.00 \pm  0.24$ & 5-10 \\
   PKS 0745-191\tablenotemark{a} &          \nodata & $1.496 \pm  0.13$  &  2.0-6.0  &  &           \nodata & \nodata & \nodata \\
   Hydra A\tablenotemark{b} &  \nodata &  $1.0 \pm  0.16$  &  1.0-5.0 &  &           \nodata & \nodata & \nodata \\
    Zw 3146\tablenotemark{c} &  $3.4 \pm  0.18$  & \nodata &  2.1-8.0 &  &           \nodata & \nodata & \nodata \\
   A1068\tablenotemark{d} &    $1.7 \pm  0.18$  & \nodata &  1.5-4.0 &  &           \nodata & \nodata & \nodata \\
   A1361             & $-0.15 \pm  0.06$ & $-0.07 \pm  0.06$ &  2.0-13.1 &  & $-0.04 \pm  0.03$ & $-0.03 \pm  0.03$ & 5-10 \\
                     & & & & & $-0.06 \pm  0.06$ & $-0.00 \pm  0.07$ & 5-20 \\
   A1413             & $-0.29 \pm  0.15$ & $-0.22 \pm  0.21$ &  1.7-9.6 &  & $-0.05 \pm  0.09$ & $-0.03 \pm  0.10$ & 5-10 \\
                     & & & & & $-0.23 \pm  0.18$ & $-0.19 \pm  0.21$ & 5-20 \\
   M87               & $-0.27 \pm  0.08$ & $-0.20 \pm  0.07$ & 11.5-96.6 &  & $-0.15 \pm  0.43$ & $-0.16 \pm  0.42$ & 5-10 \\
   HCG 62            & $-0.33 \pm  0.04$ & $-0.24 \pm  0.05$ &  2.3-42.1 &  & $-0.07 \pm  0.11$ & $-0.14 \pm  0.13$ & 5-10 \\
   A1650             & $-0.14 \pm  0.11$ & $-0.07 \pm  0.14$ &  2.1-19.7 &  & $-0.05 \pm  0.11$ & $-0.02 \pm  0.11$ & 5-10 \\
                     & & & & & $-0.20 \pm  0.17$ & $-0.14 \pm  0.20$ & 5-20 \\
   Coma              & $-0.14 \pm  0.02$ &           \nodata &  1.7-94.4 &  & $-0.04 \pm  0.08$ &           \nodata & 5-10 \\
                     & & & & & $-0.06 \pm  0.18$ & \nodata & 5-20 \\
   A1795             & $ 0.50 \pm  0.10$ & $ 0.68 \pm  0.11$ &  2.1-9.1 &  & $ 0.05 \pm  0.08$ & $ 0.14 \pm  0.08$ & 5-10 \\
                     & & & & & $ 0.07 \pm  0.18$ & $ 0.14 \pm  0.20$ & 5-20 \\
    A1835\tablenotemark{e} &    $4.5 \pm  0.18$  & \nodata &  2.0-7.0 &  &           \nodata & \nodata & \nodata \\
   A1991             & $-0.35 \pm  0.06$ & $-0.23 \pm  0.08$ &  2.6-16.9 &  & $-0.11 \pm  0.05$ & $-0.08 \pm  0.06$ & 5-10 \\
                     & & & & & $-0.13 \pm  0.10$ & $-0.02 \pm  0.12$ & 5-20 \\
   MS 1455.0+2232    & $-0.06 \pm  0.23$ & $ 0.02 \pm  0.25$ &  1.5-5.8 &  & \nodata & \nodata & \nodata \\
   RXC J1504.1-0248  & $ 0.94 \pm  0.17$ & $ 1.03 \pm  0.16$ &  1.7-5.8 &  & \nodata & \nodata & 5-10 \\
   A2029\tablenotemark{a} &          \nodata  & $-0.269 \pm  0.014$ &  3.0-10.0 &  &           \nodata & \nodata & \nodata \\
   A2052\tablenotemark{a} &           \nodata & $-0.069 \pm  0.021$ &  5.0-35.0 &  &           \nodata & \nodata & \nodata \\
   A2065             & $-0.38 \pm  0.08$ & $-0.31 \pm  0.08$ &  1.4-29.2 &  & $ 0.01 \pm  0.10$ & $ 0.04 \pm  0.10$ & 5-10 \\
                     & & & & & $-0.20 \pm  0.18$ & $-0.11 \pm  0.18$ & 5-20 \\
   RX J1532.8+3021   & $ 1.04 \pm  0.40$ & $ 1.56 \pm  0.53$ &  2.0-5.4 &  & \nodata & \nodata & \nodata \\
   A2218             & $-0.41 \pm  0.11$ & $-0.40 \pm  0.12$ &  1.4-8.7 &  & $ 0.02 \pm  0.10$ & $ 0.02 \pm  0.10$ & 5-10 \\
                     & & & & & $-0.24 \pm  0.11$ & $-0.25 \pm  0.11$ & 5-20 \\
   Hercules A        & $ 0.34 \pm  0.25$ &           \nodata &  1.6-8.8 &  & $ 0.13 \pm  0.11$ &           \nodata & 5-10 \\
                     & & & & & $ 0.20 \pm  0.32$ & \nodata & 5-20 \\
   A2244             & $-0.20 \pm  0.12$ & $-0.02 \pm  0.13$ &  1.9-18.1 &  & $-0.12 \pm  0.09$ & $-0.05 \pm  0.10$ & 5-10 \\
                     & & & & & $-0.28 \pm  0.17$ & $-0.17 \pm  0.17$ & 5-20 \\
   NGC 6338          &           \nodata & $-0.12 \pm  0.01$ &  1.6-102.2 &  &           \nodata & $-0.04 \pm  0.03$ & 5-10 \\
                       & & & & & \nodata & $-0.14 \pm  0.09$ & 5-20 \\
   RX J1720.2+2637   & $ 0.28 \pm  0.14$ & $ 0.46 \pm  0.17$ &  1.5-11.7 &  & $ 0.14 \pm  0.10$ & $ 0.16 \pm  0.11$ & 5-10 \\
                     & & & & & $ 0.27 \pm  0.18$ & $ 0.41 \pm  0.19$ & 5-20 \\
   MACS J1720.2+3536 & $ 0.84 \pm  0.85$ & $ 1.06 \pm  0.88$ &  1.6-2.2 &  & \nodata & \nodata & \nodata \\
   A2261             & \phn$-0.42 \pm  0.19$ & $-0.35 \pm  0.19$\phn &  1.4-7.4 &  & $-0.10 \pm  0.10$ & $-0.10 \pm  0.10$ & 5-10 \\
                     & & & & & $-0.31 \pm  0.17$ & $-0.26 \pm  0.17$ & 5-20 \\
   A2319             &           \nodata & $-0.13 \pm  0.05$ &  1.3-18.0 &  &           \nodata & $-0.06 \pm  0.07$ & 5-10 \\
                       & & & & & \nodata & $-0.14 \pm  0.13$ & 5-20 \\
   A2390             & $ 1.35 \pm  0.67$ & $ 1.00 \pm  0.76$ &  1.9-4.0 &  & $ 0.77 \pm  0.17$ & $ 0.67 \pm  0.17$ & 5-10 \\
                     & & & & & $ 1.41 \pm  0.54$ & $ 1.39 \pm  0.56$ & 5-20 \\
   A2409             & $-0.34 \pm  0.26$ & $-0.16 \pm  0.28$ &  1.5-6.3 &  & $-0.12 \pm  0.17$ & $-0.08 \pm  0.17$ & 5-10 \\
   A2597             & $ 0.55 \pm  0.14$ & $ 0.72 \pm  0.14$ &  2.1-8.6 &  & $ 0.23 \pm  0.10$ & $ 0.31 \pm  0.10$ & 5-10 \\
                     & & & & & $ 0.23 \pm  0.32$ & $ 0.29 \pm  0.32$ & 5-20 \\
   A2626             & $-0.09 \pm  0.03$ & $-0.02 \pm  0.04$ &  1.1-18.8 &  & $-0.01 \pm  0.07$ & $-0.01 \pm  0.07$ & 5-10 \\
                     & & & & & $-0.03 \pm  0.13$ & $ 0.03 \pm  0.13$ & 5-20 \\
   A2657             & $-0.14 \pm  0.06$ & $-0.08 \pm  0.07$ &  1.4-9.5 &  & $-0.02 \pm  0.15$ & $ 0.01 \pm  0.15$ & 5-10 \\
                     & & & & & $-0.01 \pm  0.39$ & $ 0.10 \pm  0.39$ & 5-20 \\
   A2670             &           \nodata & $-0.26 \pm  0.03$ &  1.4-32.2 &  &           \nodata & $-0.08 \pm  0.03$ & 5-10 \\
                       & & & & & \nodata & $-0.18 \pm  0.06$ & 5-20 \\
\enddata
\tablenotetext{a}{Gradient taken from \citet{mcna92}}
\tablenotetext{b}{Gradient taken from \citet{mcna95}. We have adopted the average $G(U-I)$ error for our sample.}
\tablenotetext{c}{Gradient taken from unpublished data (B.~R.~McNamara, private communication). We have adopted the average $G(U-R)$ error for our sample.}
\tablenotetext{d}{Gradient taken from \citet{mcna92}. We have adopted the average $G(U-R)$ error for our sample.}
\tablenotetext{e}{Gradient taken from \citet{mcna06}. We have adopted the average $G(U-R)$ error for our sample.}
\end{deluxetable*}

The $U-R$ and $U-I$ color gradients are very similar, with the $U-I$ color gradients being on average slightly more positive (due to a lower contribution from the star formation to the flux in the $I$ band). To illustrate this, we plot in Figure \ref{F:GU_R_vs_GU_I} the two gradients against one another. Since the errors in the two gradients are correlated through the errors in the $U$-band surface-brightness profiles, we perform a linear regression using the BCES least-squares method of \citet{akri96}, which properly accounts for correlated errors. The resulting fit (with 1-$\sigma$ errors) is:
\begin{equation}
 G(U-I) = (0.090 \pm 0.015) + (1.01 \pm 0.14)G(U-R).
\end{equation}
The $U-I$ color gradients are therefore consistent with a simple offset from the $U-R$ color gradients. We note that there are no objects for which the difference between the two gradients is large enough ($\gtrsim 2 \sigma$) to indicate significant contamination from emission lines in the $R$ or $I$ images. 

The gradients discussed above do not by themselves give an indication of the radial extent of the star formation. For two objects with the same gradient, one may have star formation over a much larger physical radius. Therefore, it is useful to derive the change in color between two physical radii. Objects with star-forming regions that are physically smaller will tend to have smaller color changes than those objects with more extended star formation. To derive the color changes, denoted $\Delta(U-R)$ and $\Delta(U-I)$, we use simple linear interpolation to estimate the colors at 5, 10, and 20 kpc. The $\Delta(U-R)$ color changes are then:
\begin{equation}
 \Delta(U-R) = (U-R)_{\rm outer} - (U-R)_{\rm inner} 
\end{equation}
Table~\ref{T:gradients} lists the $\Delta$ colors for the optical sample.

\begin{figure*}
\includegraphics[width=1.0\textwidth]{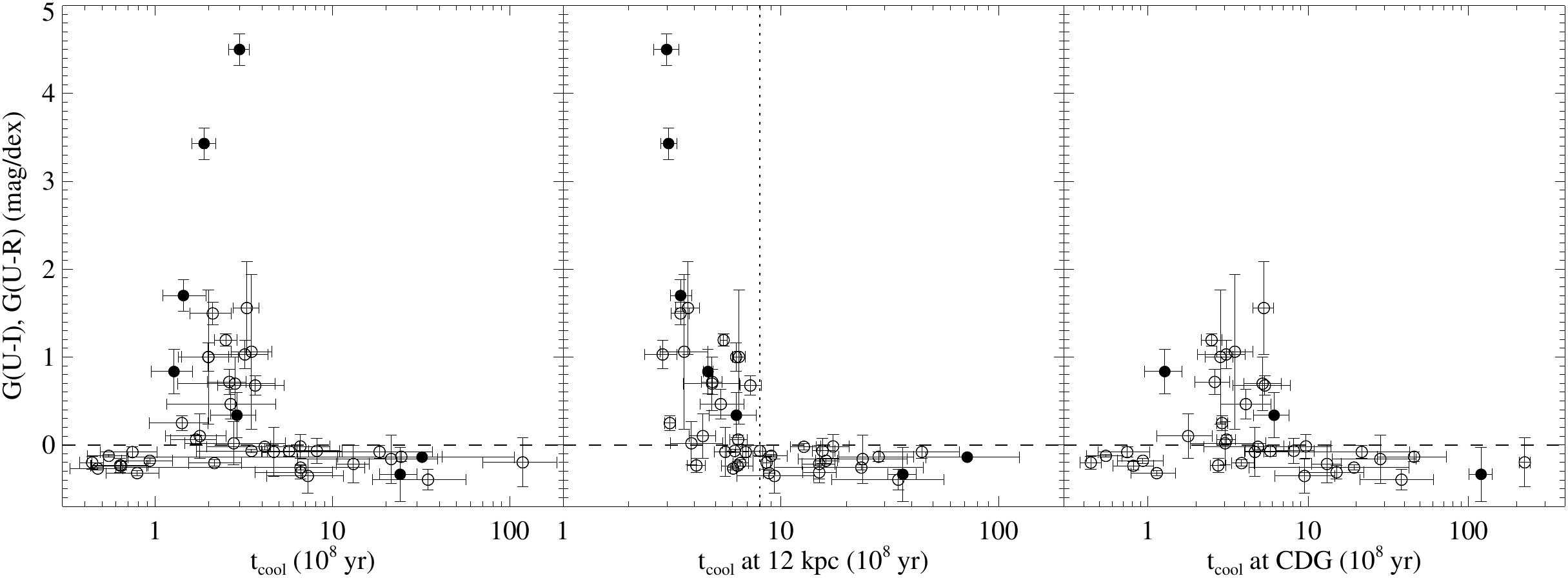}
\caption{Color gradient versus the central cooling time (\textit{left panel}), the cooling time at a radius of 12 kpc (\textit{middle panel}), and the cooling time at the location of the CDG (\textit{right panel}). $U-I$ gradients (\textit{empty symbols}) are used when available; $U-R$ gradients (\textit{filled symbols}) are used when $U-I$ gradients are unavailable. The horizontal, dashed line marks the division between systems with positive color gradients (blue, star-forming systems) and those with negative color gradients (red systems). The vertical line in the middle panel marks $t_{\rm cool}(\mbox{12 kpc}) = 8\times10^8$ yr.\label{F:Grad_vs_tcool}}
\end{figure*}

The detectability of blue emission depends on several factors, such as the signal-to-noise ratio of the images, the seeing or PSF size (as we exclude unresolved emission), and the redshift of the source. Since this PSF size limits our spatial resolution, an obvious bias is that we are sensitive to blue emission that extends over smaller physical radii in lower-redshift systems than in higher-redshift systems. This bias could be important if the red systems are at systematically higher redshift than the blue systems, such that extended blue emission could be missed preferentially in the red systems. However, the mean redshift of the blue systems ($<z>=0.159$) is approximately twice that of the red systems ($<z>=0.073$). The lower mean redshift of the red systems is likely due to X-ray selection effects, because, as we demonstrate in Section \ref{S:SF_cooling}, the red systems often lack short central cooling times and cool cores, making them more difficult to image in X-rays at a given redshift.

\section{Discussion}

\subsection{The Relation between Star Formation and Cooling Time}\label{S:SF_cooling}
The cooling time of the ICM is a critical measure of its thermal state. If the cooling time is sufficiently short (on the order of the age of the system or less), significant condensation of gas should occur unless heating balances the cooling entirely. In this section, we investigate whether the cooling times we derived from \textit{Chandra} data are related to the presence of excess blue emission in the CDG. 

To this end, in Figure \ref{F:Grad_vs_tcool} we plot the optical color gradients of the CDG against the central cooling times of the cluster's ICM. We plot the $U-I$ color gradient when available; otherwise, we plot the $U-R$ gradient, as they are almost equivalent (see Section \ref{S:Color_gradients}). The left panel shows the emission-weighted central cooling time derived as close to the core as our data allow us to achieve (see Table~\ref{T:ICM_central}). Since the systems in our sample vary greatly in redshift, we plot in the middle panel the cooling time at a single physical radius of 12 kpc (see Table~\ref{T:ICM_12kpc}). At 12 kpc, the very short cooling times (those below $\sim 2 \times 10^{8}$ yr) measured at the center disappear, demonstrating that their short cooling times relative to the rest of the sample are due to resolution effects. 

\begin{figure}
\plotone{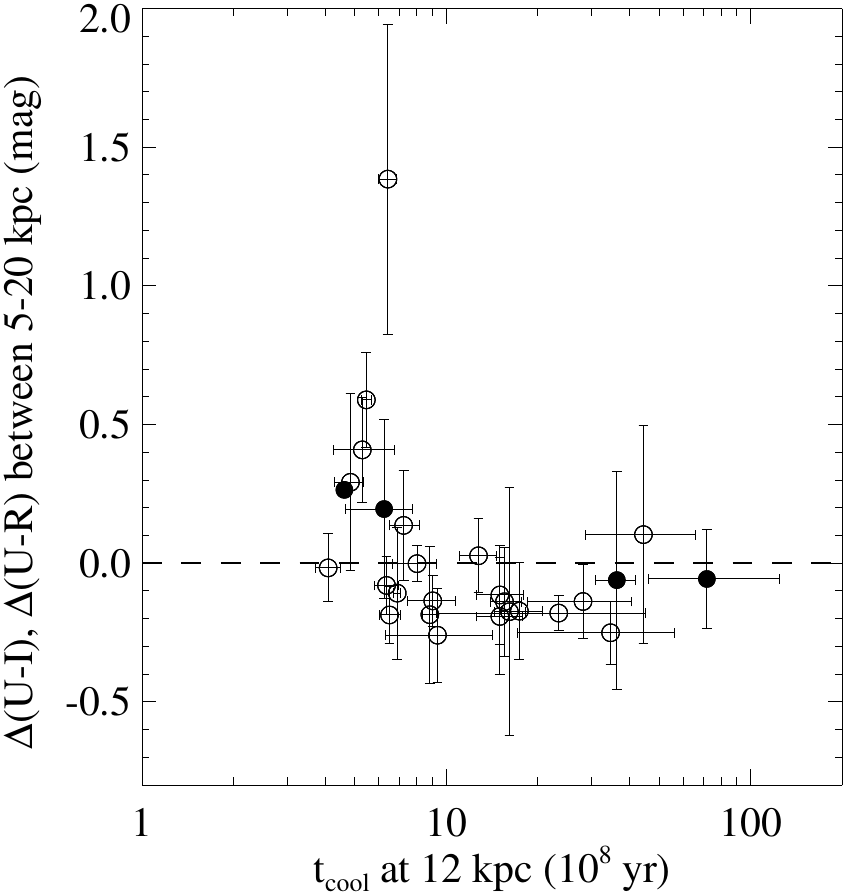}
\caption{$\Delta(U-R)$ and $\Delta(U-I)$ between 5-20 kpc versus the cooling time at 12 kpc. Symbols are the same as those in Figure \ref{F:Grad_vs_tcool}.\label{F:D_vs_tcool}}
\end{figure}

For comparison, we also plot the $\Delta$ colors against the cooling times in Figure \ref{F:D_vs_tcool}. In this figure, the overall relationship between cooling time and the presence of star formation is unchanged from that shown in Figure \ref{F:Grad_vs_tcool}. If the star formation rate is governed by the local value of the cooling time at each point throughout the ICM, then we should expect the $\Delta$ colors to increase smoothly as the cooling time at 12 kpc decreases. The abrupt turn on of star formation at $t_{\rm cool} \simeq 4$--$8\times 10^8$ yr in Figure \ref{F:D_vs_tcool} does not support such behavior.

It is apparent from Figure \ref{F:Grad_vs_tcool} that positive gradients, which are indicative of active star formation, occur only in objects with central ($r \lesssim 12$ kpc) cooling times below $\sim 7$--$8 \times 10^8$ yr. Such a direct relationship between star formation and the local state of the intracluster gas strongly implicates the gas as the cause of the star formation. The simplest possibility is that cooling gas fuels the star formation directly. Alternatively, dense gas might have fueled an AGN outburst that triggered the star formation. However, it is also apparent that the presence of a short central cooling time does not guarantee the presence of a positive color gradient, as several of the objects with cooling times below $\sim 8 \times 10^8$ yr do not have blue cores (although the majority do). 

\begin{figure}
\includegraphics[width=0.45\textwidth]{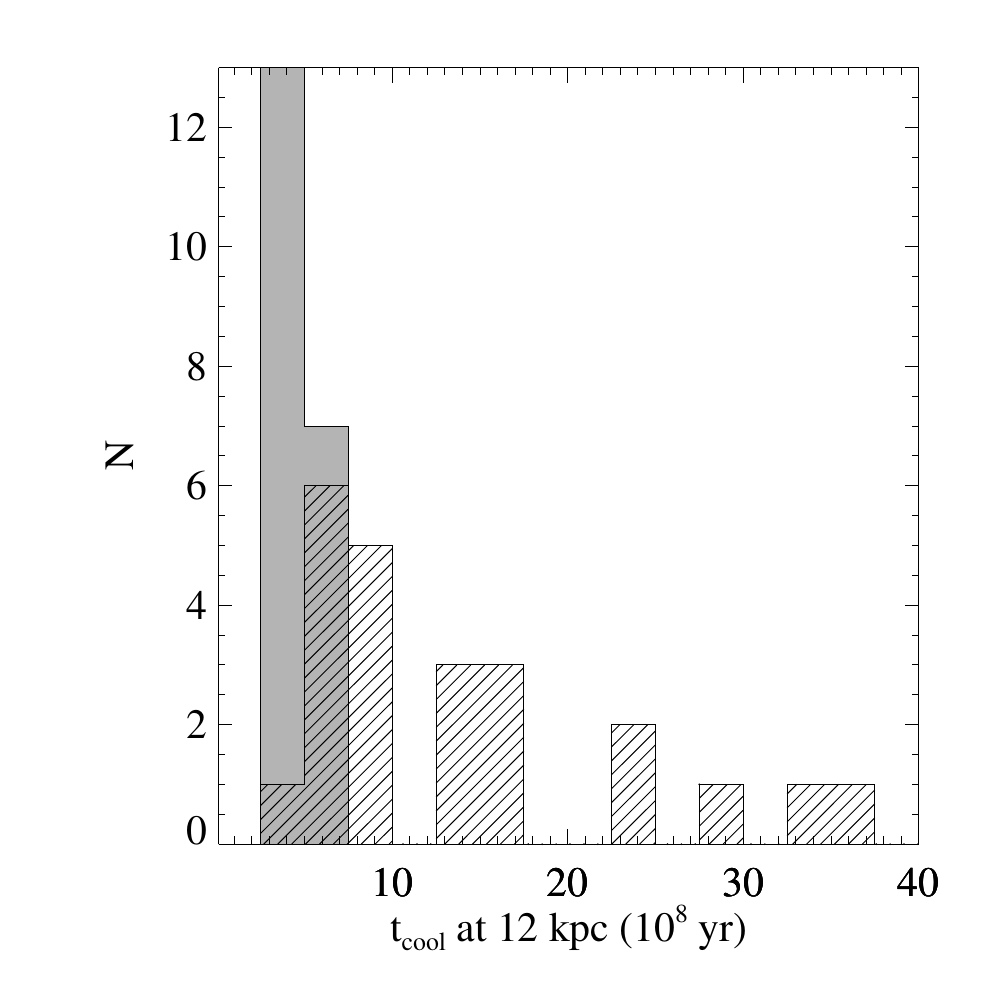}
\caption{Histogram of the cooling times at 12 kpc for objects with positive color gradients (blue systems, \textit{gray region}) and negative gradients (red systems, \textit{hatched region}). Three red systems with cooling times greater than $4 \times 10^9$ yr are not shown. \label{F:tcool_hist}}
\end{figure}
To emphasize that the systems with blue central colors are different on average from those with red colors, we show in Figure \ref{F:tcool_hist} histograms of the cooling times at 12 kpc for the objects in our optical sample with cooling times below $4 \times 10^9$ yr. We have divided the sample into those systems with positive color gradients (blue systems) and those with negative gradients (red systems). We include MS 1455.0+2232, which has a positive $U-I$ gradient and a negative $U-R$ gradient, in the blue sample. Compared with the red systems, the blue systems clearly prefer shorter cooling times.  A Kolmogorov-Smirnov (K-S) test gives a probability of just $1.3 \times 10^{-7}$ that the red and blue samples are drawn randomly from the same parent distribution of cooling times. This result is a strong indicator that the cooling flow and star formation in the CDG are connected.

The large disparity between the mass of gas within the cooling regions of the cluster and the mass of cold gas and new stars, coupled with the sharply declining cooling times with decreasing radius, precludes the possibility that the central cooling time is significantly affected by the cold gas. Thus, the tight relationship between cooling times and color gradients make it implausible that the stars are formed from infalling cold gas.

One possibility is that thermal instabilities in the hot, cooling gas are driving the star formation.  Under the right conditions, the motion of a cooling blob of gas will become unstable, leading eventually to catastrophic cooling. These conditions are met roughly when the growth rate of the instabilities due to cooling exceeds the counteracting rate of any damping, such as that from viscous damping (which acts to slow the blob's radial oscillations) or thermal conduction (which acts to reduce the blob's temperature contrast with its surroundings). The stability condition in any region can be determined from the X-ray data, and a detailed treatment of this scenario will be presented in a forthcoming paper.
\begin{figure*}
\includegraphics[width=1.0\textwidth]{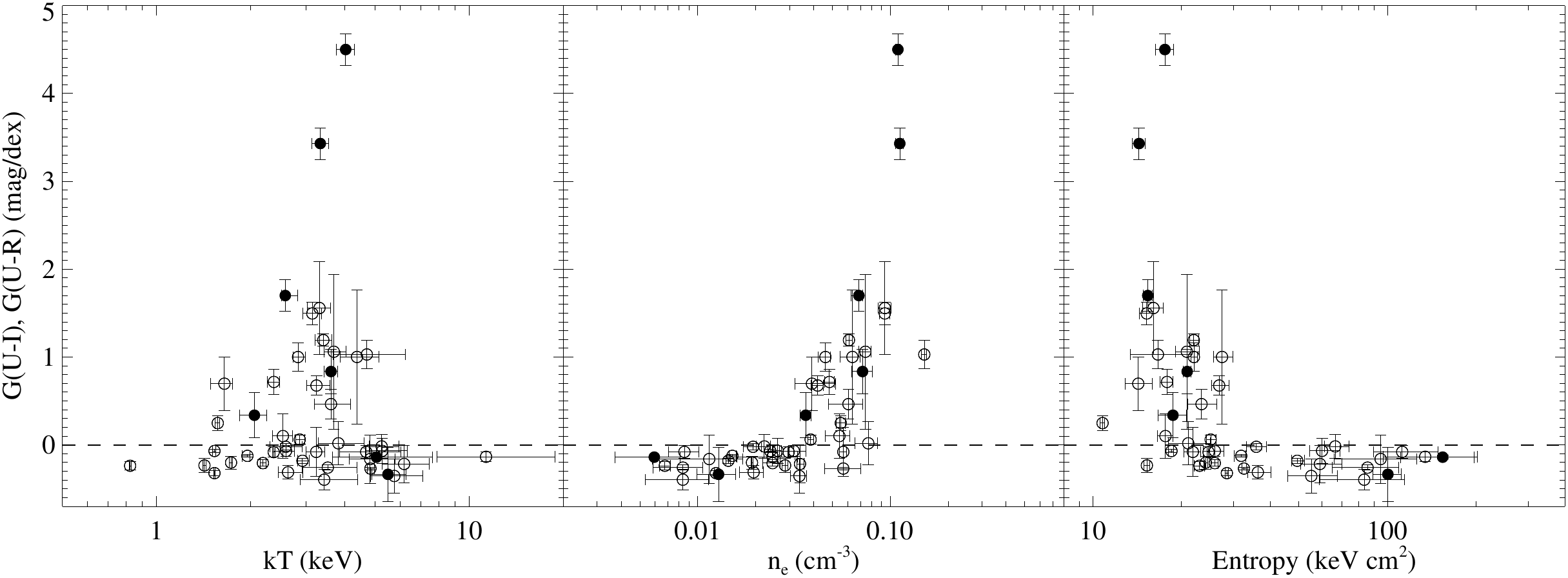}
\caption{Color gradient versus the temperature, density, and entropy at 12 kpc. Symbols and the dashed line are the same as those in Figure \ref{F:Grad_vs_tcool}. \label{F:Grad_vs_kT_pres}}
\end{figure*}

\subsection{Star Formation and Other ICM and Galaxy Properties}\label{S:other_props}
We also investigated the dependence of the presence of excess central blue emission on the temperature, density, and entropy of the ICM. We plot the CDG color gradients against each of these properties, derived at a single physical radius of 12 kpc, in Figure \ref{F:Grad_vs_kT_pres}. The temperature shows no clear relation to the color gradient. However, it appears that high densities ($\gtrsim 0.03$ cm$^{-3}$) are required for blue colors, in agreement with the cooling time findings discussed earlier (since $t_{\rm cool} \propto n_e^{-1}$). 

Additionally, we find that positive gradients occur only in objects in our sample whose entropies at 12 kpc are $\lesssim 30$ keV cm$^2$ (see the right-hand panel of Figure \ref{F:Grad_vs_kT_pres}). The range in entropy levels seen in this plot for the blue objects is consistent with that found by  \citet{dona06} for a sample of 9 cooling-flow clusters (all of which are shared with our sample). \citet{voit05a} show that such entropy levels of $\sim 10$--30 keV cm$^2$ are consistent with those expected from the effects of AGN heating on a simple cooling entropy profile, implying an indirect connection between feedback and star formation.

The great majority of CDGs host a radio source \citep{burn90}. Therefore, it is useful to examine whether the presence of a radio source (and hence an AGN) is related to the presence of star formation. To identify AGNs, we have searched available radio catalogs (e.g., the NVSS and FIRST catalogs) for radio sources associated with the CDGs in our sample. We find detections at 1400 MHz for 42 of our 46 CDGs. The four sources that lack detections are A1413, A2218, AWM 7, and A2261, all of which have negative central color gradients (and consequently no evidence of recent star formation) and long central cooling times. However, the remaining 21 systems with negative gradients have radio sources and, therefore, the presence of an AGN does not appear to depend stongly on the presence of star formation in the CDG (and vice versa). A detailed study of the  relation between the radio-source and star-formation properties is beyond the scope of this paper and will presented in a future paper.

\begin{figure}
\includegraphics[width=8.5cm]{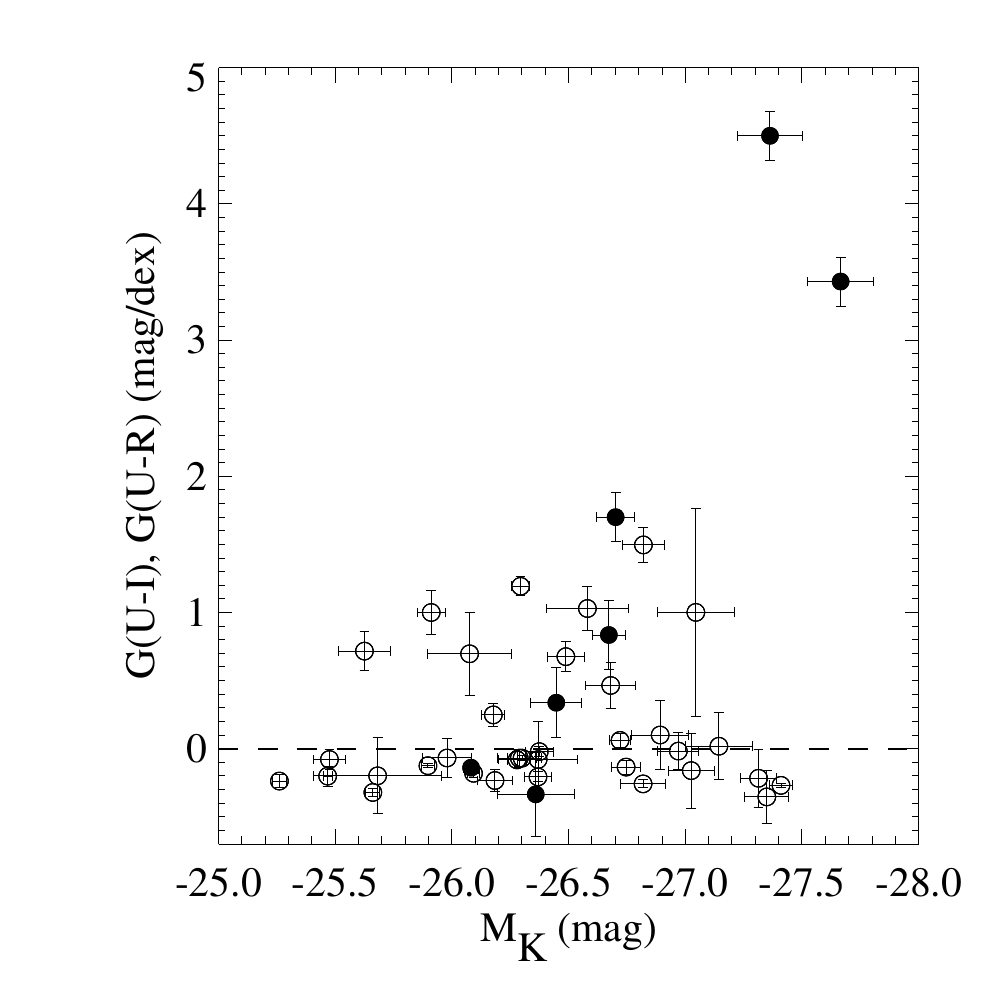}
\caption{Color gradient versus the total $K$-band absolute magnitude of the galaxy.  Symbols and the dashed line are the same as those in Figure \ref{F:Grad_vs_tcool}. \label{F:G_vs_M_K}}
\end{figure}
Lastly, we investigated whether the color gradient has some dependence on the total luminosity of the galaxy. We plot in Figure \ref{F:G_vs_M_K} the color gradient versus the total $K$-band luminosity of the CDG. Apparent, total $K$-band magnitudes were taken from the Two Micron All Sky Survey (2MASS) catalog.\footnote{See \url{http://www.ipac.caltech.edu/2mass/}.} The apparent magnitudes were corrected for Galactic extinction with the values of \citet{schl98} and corrected for redshift ($K$-corrected) and evolution using the corrections of \citet{pogg97}.  Lastly, the magnitudes were converted to absolute magnitudes using our assumed cosmology and the redshifts listed in Table \ref{T:sample} (which also gives the resulting absolute magnitudes).  It is clear from Figure \ref{F:G_vs_M_K} that the color gradient shows no clear relation to the total luminosity of the galaxy; positive gradients are found across almost the full range of luminosities. Additionally, a clear dichotomy between quiescent red galaxies and blue star-forming galaxies with steep gradients is evident.

\subsection{Cooling, Star-Formation, and AGN-Heating Timescales}
In the simple AGN feedback model, the timescales over which cooling, star formation, and heating occur should be related. The presence of shocks, in particular, suggest that AGN heating is intermittent \citep[e.g.,][]{nuls05a,nuls05b,wils06,fabi06,form07}. In such a model, the AGN must create cavities frequently enough to prevent large amounts of net cooling and to maintain the gas at the observed temperatures.  This condition can only be met when the average time between outbursts is less than the average central cooling time. Additionally, cooling must occur over a long enough period for significant cool gas to accumulate and star formation to occur, after which the signatures of star formation may persist even though cooling has ceased. The probable interrelation of the various timescales will lead to temporally induced scatter whenever quantities that depend on these timescales are compared. 

\begin{figure}
\plotone{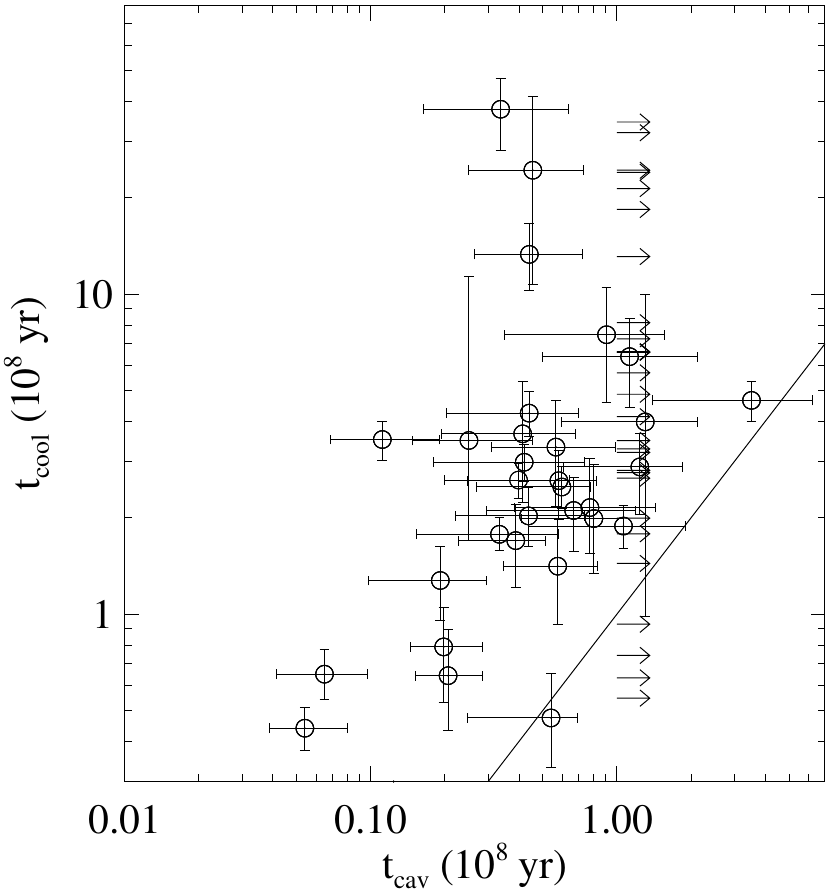}
\caption{The central cooling time versus the average cavity age from \citet{raff06}. The errors in age reflect the range in the three age estimates (see text for details). Lower limits are shown for systems without visible cavities under the assumption that any cavities must have disrupted (or faded below detection limits) and are therefore likely older than $10^8$ yr.  The line denotes equality between the two times. \label{F:tcav_tcool}}
\end{figure}

We plot in Figure \ref{F:tcav_tcool} the current central cooling time against the average cavity age calculated as the average of three standard age estimates: the sound speed age, the buoyancy age, and refill age \citep[for details, see][]{raff06}. For systems with two cavities, an average is taken across both cavities. For systems with multiple generations of cavities (Perseus and Hydra A), the average age is the average difference in ages between the inner and outer sets of cavities (i.e.\ the time between outbursts). We adopt the range in age from the three estimates as the error. For systems unique to the \citet{raff06} sample, cooling times were derived following the procedure described in Section \ref{S:X-ray_analysis}.

We note that cavities are rarely seen with ages greater than $\sim 10^8$ yr \citep{mcna07}, perhaps because they disrupt on that timescale or fade into the background as they rise. Therefore, systems without visible cavities would likely lie at  inter-outburst times (cavity ages) above $10^8$ yr in Figure \ref{F:tcav_tcool}. In keeping with this conclusion, such systems are shown in Fig \ref{F:tcav_tcool} as lower limits. For three systems in our sample (A496, A1991, and NGC 6338), this limit exceeds the central cooling time by a significant factor, implying that feedback cannot operate effectively and significant cooling should occur.  In A1991, \citet{mcna89,mcna92} found evidence for recent star formation (while we do not find a positive color gradient in A1991, our images of this system were affected by poor seeing and scattered light, and hence the excess blue emission was likely missed). The two remaining systems do not show evidence of recent star formation. It is possible that these systems do have cavities, but that they are below our detection limit (indeed, the existing \textit{Chandra} image of A496 shows some evidence of faint cavities).

Figure \ref{F:tcav_tcool} demonstrates that the average cavity age in all systems with detected cavities is less than the current cooling time, a result that is consistent with feedback. We note that there is evidence of a trend between the two times, as expected if systems with shorter cooling times need more frequent outbursts to prevent the cooling of large amounts of gas. However, the trend is very weak, as systems with similar cavity ages have cooling times that differ by up to a factor of $\sim 100,$ possibly because the true average time between outbursts is not known for the majority of systems in our sample (for systems without multiple generations of cavities, the current cavity age is only a lower limit on the average time between outbursts). The central cooling time may also vary over time as the ICM is heated and cools, inducing additional scatter in Figure \ref{F:tcav_tcool}.

The timescales relevant for star formation are the time for cool gas to form stars and the time for star formation to fade below detection limits. The former timescale is not well known, but is believed to be on the order of $10^7$ yr \citep[e.g.,][]{mous06}. The timescale for star formation to fade may be roughly estimated for our sample using the $\Delta$ colors given in Table \ref{T:gradients} and the approximately power-law relation between the color and age of a young stellar population. For a burst of star formation with a metallicity one-half solar, the stellar population synthesis models of \citet{bruz03} predict that the $U-I$ color fades at a rate of $\Delta (U-I)/\Delta \log t \approx 1.2$ mag dex$^{-1}$. If we assume that the average age for the star forming regions in our sample is $\sim 10^8$ yr, a $\Delta(U-I)$ color of 0.3 mag (typical of the blue objects in our sample) would fade to $\Delta(U-I) \sim -0.2$ (typical of the red objects) in a few $10^8$ yr. Alternatively, the mean change across the sample in $U-I$ or $U-R$ color calculated between the inner and outer radii of star formation (instead of between 5-10 kpc or 5-20 kpc as was done for the $\Delta$ colors) is $\approx 0.52,$ resulting in the same fading time. Consequently, detectable emission from star formation should typically persist for a few $10^8$ yr after cooling ceases. 

The total process of gas condensation, star formation, and the fading of the UV emission should therefore require $\sim 5 \times 10^8$--$10^9$ yr. This timescale is longer than the typical time between outbursts or the typical central cooling time of the systems in our sample, thus the colors that we observe could represent the properties of star-forming populations averaged over several heating events. Additionally, the cooling time at 12 kpc, which is comparable to this total timescale, could roughly trace the state of the ICM averaged over timescales long enough to be typical of the conditions under which the stars we see now were accumulated. This scenario might explain the strong dependence of star formation on the cooling time at 12 kpc seen in Figure \ref{F:Grad_vs_tcool}.

\subsection{Star Formation and the CDG's Location}
\begin{figure}
\includegraphics[width=0.45\textwidth]{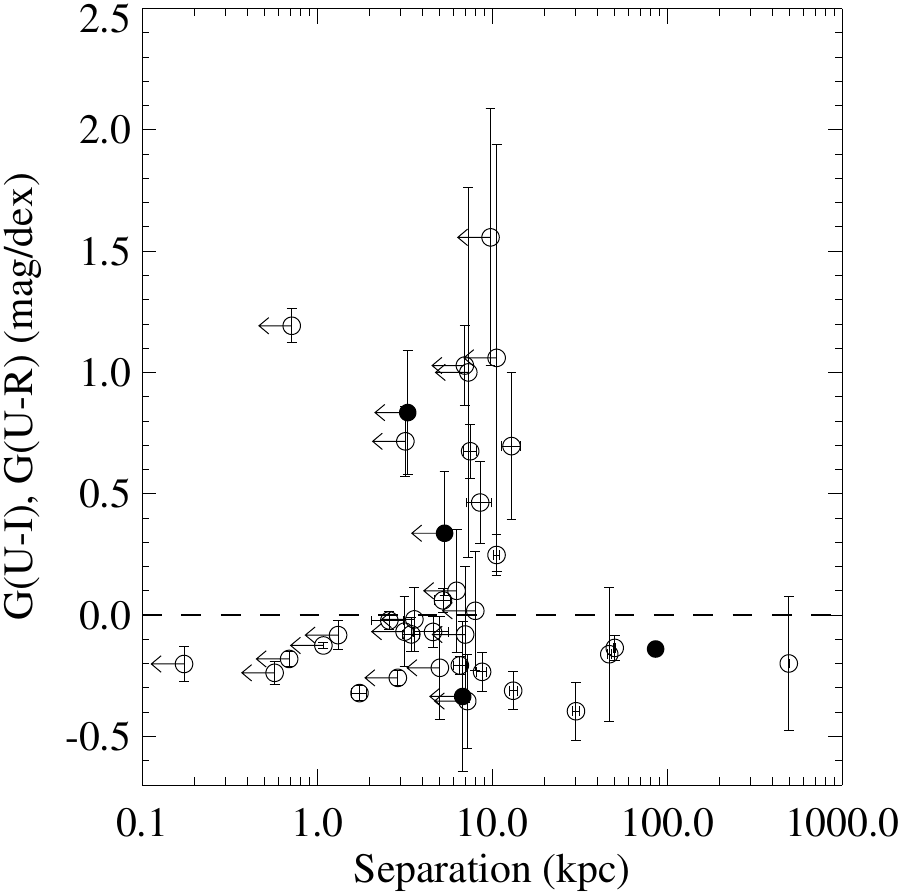}
\caption{Color gradient versus the cooling time at the projected radius of the CDG's core. Symbols and the dashed line are the same as those in Figure \ref{F:Grad_vs_tcool}. \label{F:G_separation}}
\end{figure}

It is quite common to find the CDG offset 100 kpc or more in projection from the X-ray core \citep[e.g.,][]{pate06,edwa07}. If the CDG does not lie at the core of the cooling flow where cooling times are short, but instead resides at larger radii where the cooling times are relatively long, we should expect little active star formation if it is fueled by the cooling ICM.  To investigate this possibility, we plot in Figure \ref{F:G_separation} the color gradient against the projected physical separation between the CDG's core and the cluster's core.  In the right panel of Figure \ref{F:Grad_vs_tcool} we plot the color gradient against the cooling time of the ICM at the projected location of the CDG's core. In neither plot do the red objects segregate to large separations or long cooling times (although those objects with large separations tend to have long cooling times). However, the blue objects all have small ($\lesssim 20$ kpc) projected separations. We note that projection could result in measured separations that are much smaller than the true ones. However, on average, such effects should be relatively small.  

In summary, although a small (projected) separation between the X-ray cusp and the galaxy does not guarantee a blue core, it appears to be a necessary condition. This result agrees with the findings of \citet{edwa07}, who find that all the CDGs in their sample with strong optical line emission (indicative of an ionizing source, such as young stars or an AGN, near cool gas) lie within 50 kpc of the cluster's X-ray core \citep[see also][]{craw99}. Therefore, the CDG's location relative to  the cluster's core is critical to the presence of star formation, a finding which supports the hypothesis that the cooling ICM, which preferentially cools in the core, fuels the star formation.

\subsection{The Suppression of Star Formation by AGN Feedback}\label{S:heat_cool_cycle}
The lack of excess blue emission in some objects with short cooling times may be due to several factors. In general, two situations could apply: either active star formation is present, but the excess blue emission is obscured, or no active star formation is present. In the former case, dust obscuration is the most likely cause; however, as we argue below, dust is unlikely to obscure all signs of significant star formation. In the latter case, where no active star formation is occurring, some mechanism for suppressing star formation must be present, since cooling times are short and significant condensation from the ICM onto the CDG should be occurring. An obvious (but perhaps not the only) mechanism for the suppression of cooling and star formation in these systems is AGN feedback. In this section, we examine the red systems with short cooling times and discuss the possibility that dust obscuration or suppression by AGN feedback could explain their properties.

Dust is often associated with cold gas and star formation and is common in the cores of CDGs \citep[e.g.,][]{lain03}. Dust will preferentially scatter and absorb short-wavelength emission, resulting in observed colors that are redder than the intrinsic ones. However, dust in CDGs is generally observed to be patchy or filamentary \citep[e.g.,][]{lain03}, not spread  smoothly across the galaxy in significant quantities \citep[see however][]{silv96}. Therefore, it is unlikely that dust would obscure the entire star forming region if star formation in all of the objects with short cooling times is similar. New far-infrared measurements of star formation in cooling flows obtained with the \textit{Spitzer} observatory roughly agree with optical and near-UV rates \citep{odea08}. Furthermore, for the 17 systems in our sample that are included in recent infrared studies \citep{egam06,odea08,quil08}, the 8 systems that show infrared evidence of star formation also have blue cores, implying that star formation in these systems is not heavily obscured. Of the remaining 9 systems without infrared-detected star formation, most (7) have red cores. The remaining two, A1991 and MS 1455.0+2232, have only weakly blue cores. Therefore, near-UV and infrared data appear to have similar sensitivity to the presence of star formation in these systems, and we are unlikely to have missed significant star formation in the red systems.

If instead star formation is suppressed in some systems with short cooling times, AGN feedback may be the primary cause. As noted in Section \ref{S:other_props}, 42 of the 46 systems in our sample have detections at 1400 MHz, and all of the systems with short cooling times have detections. Therefore, AGNs appear to be almost ubiquitous in the cores of cooling flows. Additionally, AGN feedback at the cores of clusters has been shown to be energetically capable of completely quenching cooling from the ICM in many systems \citep{birz04,raff06,dunn06}. Some systems with recent AGN outbursts seem to be in a heating stage, in which AGN feedback is supplying excess heat above that required to balance cooling losses from the ICM, whereas other systems appear to be in a cooling stage, in which AGN feedback cannot balance the entire cooling luminosity of the cluster. Simulations suggest that these systems may cycle between the two stages \citep{omma04a,ciot07}. An intriguing possibility is that the blue systems occupy different temporal locations on the cycle of cooling and heating than the red systems. The blue systems could reside in clusters in which the central AGN is not currently supplying sufficient heat to offset cooling, whereas in the red systems, the AGN prevents large amounts of cooling from occurring, despite the sometimes short cooling times. In this scenario, one might expect that those objects with large amounts of heating relative to cooling would be redder than those with insufficient heating. 

\begin{figure}
\plotone{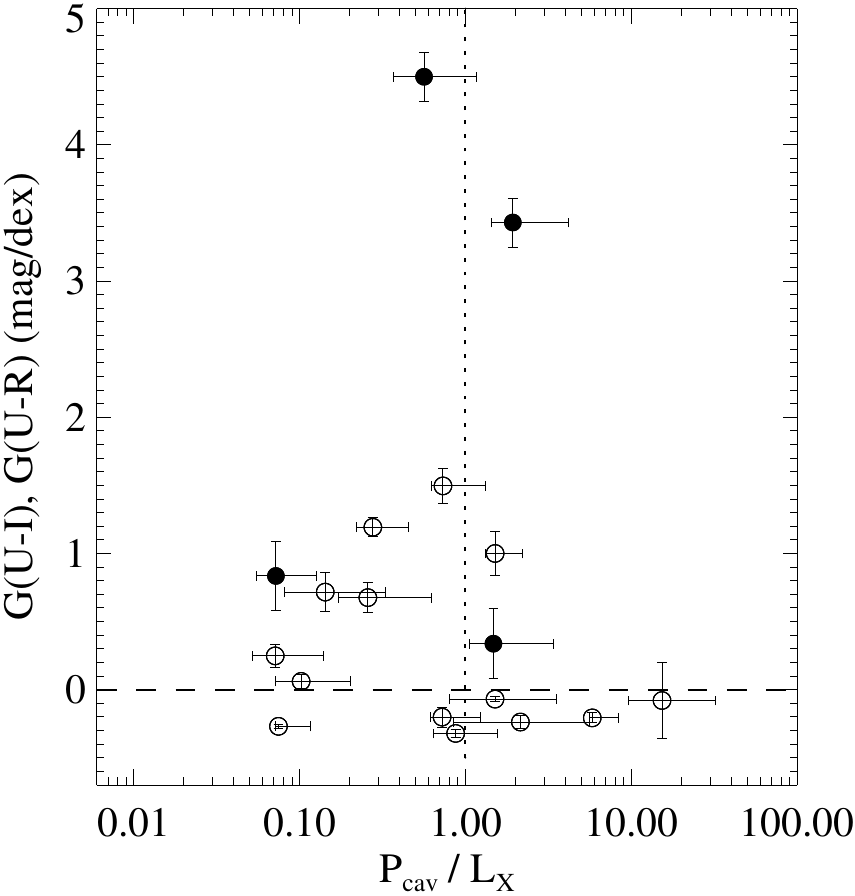}
\caption{Color gradient versus the ratio of cavity power to cooling luminosity. The vertical, dotted line divides systems with cavity power in excess of that needed to balance the cooling luminosity ($P_{\rm cav}/L_{\rm X} \gtrsim 1$) from those with insufficient cavity power ($P_{\rm cav}/L_{\rm X} \lesssim 1$) to do so. The dashed line and symbols are the same as those in Figure \ref{F:Grad_vs_tcool}.  \label{F:Grad_vs_P-L}}
\end{figure}
To test this prediction, we plot in Figure \ref{F:Grad_vs_P-L} the color gradient versus the ratio of cavity power to ICM luminosity within the cooling radius, $P_{\rm cav}/L_{\rm X},$ where $P_{\rm cav}$ and $L_{\rm X}$ were taken from \citet{raff06}. $P_{\rm cav}$ was calculated as $4pV/t_{\rm cav}$, and  $L_{\rm X}$ is the bolometric luminosity of the X-ray--emitting gas within the radius inside which $t_{\rm cool} < 7.7 \times 10^9$ yr. While our sample lacks a large number of systems with a high ratio of cavity power to cooling luminosity \citep[the X-ray--cavity sample of][however, consists of roughly equal numbers of systems above and below a ratio of unity]{raff06}, it appears that systems with excess blue emission are more likely to have a low ratio of $P_{\rm cav}/L_{\rm X}$ (i.e.\ insufficient heating to balance cooling).  The average (median) ratio of $P_{\rm cav}/L_{\rm X}$ for objects with detected X-ray cavities and positive gradients is 0.65 (0.28); for objects with negative gradients, it is 3.78 (1.51). However, a K-S test does not rule out the possibility that the positive- and negative-gradient cavity samples share the same parent distribution of the ratio of cavity power to ICM luminosity (the resulting probability that they do is 0.17). A larger sample of systems with high ratios of $P_{\rm cav}/L_{\rm X}$ will be critical to test the hypothesis that AGN feedback is quenching star formation.

For comparative purposes, we estimated $P_{\rm cav}$ for the remaining systems (that lack evidence of X-ray cavities) with short cooling times (those with $t_{\rm cool}[\mbox{12 kpc}] \lesssim 8 \times 10^8$ yr) from the monochromatic 1400 MHz radio luminosities using the jet-power scaling relation of \citet{birz08}:
\begin{equation}\label{E:pcav_p1400}
\log P_{\rm cav} = (0.35\pm 0.07)\log P_{1400} + (1.85\pm 0.10),
\end{equation}
where $P_{\rm cav} $ has units of $10^{42}$ erg s$^{-1}$ and $P_{1400}$ has units of $10^{24}$ W Hz$^{-1}$. The radio fluxes for all systems were taken from the VLA FIRST \citep{beck95} or NVSS \citep{cond98} catalogs, with the exception of A2390, the flux of which was taken from \citet{owen82}. This scaling relation was calibrated using a large sample of X-ray--cavity systems in clusters (many of which are also in our sample) and should therefore be generally applicable to our sample. It should be noted that the scatter about this relation is large ($\sigma \approx 0.8$ dex), and hence the cavity power for a given system may be misestimated by a large factor. However, averages of the resulting cavity powers over many systems should be reasonably reliable. 

Of the 12 systems for which the cavity power was estimated in this way, 10 have evidence of recent star formation. We include A1991 among these system, since \citet{mcna92} found a positive $U-I$ gradient in A1991 ($G[U-I]=0.036 \pm 0.022$), and \citet{mcna89} found spectral evidence of a weak blue excess; therefore, we have probably missed the blue emission in this system (our observations were affected by scattered light and poor seeing). These 10 systems have $\left< P_{\rm cav}/L_{\rm X}\right> = 0.1$. The remaining 2 systems (A1361 and A496), with negative gradients, have $\left< P_{\rm cav}/L_{\rm X}\right> = 0.45$.

Additionally, a search of the literature \citep[e.g.,][]{craw93,roch00,hick05} reveals that, of the 32 cavity systems studied in \citet{raff06}, only 3--6 of the 16 systems with $P_{\rm cav}/L_{\rm X} > 1$ show evidence of recent star formation, whereas 10--13 of the 16 systems  with $P_{\rm cav}/L_{\rm X} < 1$ do. While the \citet{raff06} sample, which was selected through visual inspection of publicly available observations in the \textit{Chandra} Data Archive, may be biased towards systems with easily detected cavities \citep[see e.g.,][]{dieh08}, such biases should not be related to the detectability of star formation. Therefore, the tendency for systems with low ratios of  $P_{\rm cav}/L_{\rm X}$ to host recent star formation and those with high ratios to lack such star formation does not appear to be due to the bias in our sample towards systems with low ratios. 

The tendency for blue systems to have low ratios of $P_{\rm cav}/L_{\rm X}$ and red systems to have high ratios lends support to the feedback scenario outlined above, and can explain 5 of the 7 red systems in Figure \ref{F:Grad_vs_tcool} with a cooling time at 12 kpc less than $\sim 8\times 10^8$~yr: A1361, A133, HCG 62, A2052, MS 0735.6+7421, all of which have $P_{\rm cav}/L_{\rm X} \gtrsim 1$. The two remaining systems are A496 ($t_{\rm cool}[\mbox{12 kpc}] \approx 7 \times 10^8$ yr and $P_{\rm cav}/L_{\rm X} \approx 0.3$) and A2029 ($t_{\rm cool}[\mbox{12 kpc}] \approx 6 \times 10^8$ yr and $P_{\rm cav}/L_{\rm X} \approx 0.07$), which do not appear to have enough AGN heating to balance cooling and yet have short cooling times and are red. It is possible that we have underestimated the cavity powers for these systems, particularly for A496, for which we estimated $P_{\rm cav}$ using equation (\ref{E:pcav_p1400}). Additionally, \citet{clar04} found a spiral excess structure in \textit{Chandra} data at the core of A2029 that most likely indicates a recent infall or merger, which may have disrupted cooling and star formation in the core without increasing the cooling time at 12 kpc greatly.  Lastly, we may have simply caught the two systems just before star formation will occur (i.e., they are transitioning to a cooling phase). 

We note that the dividing point between blue and red systems appears to be roughly at $P_{\rm cav}/L_{\rm X} \approx 1$, implying that, if the above scenario is correct, the cavity power is a good tracer of the heat input of the AGN. Cavities may trace the total heat input well because their powers scale with those of associated heating mechanisms, such as shocks \citep[e.g.,][]{nuls05a,mcna05} and sound waves \citep{sand07}. It is also possible that cavities are the dominant heating mechanism in these systems. Nevertheless, it appears from Figure \ref{F:Grad_vs_P-L} that the assumptions that go into the calculation of the cavity power and cooling luminosity are approximately correct, or the dividing line would be shifted far from unity. The division between blue and red systems is not perfectly clean, however, as some blue systems fall somewhat above unity and some red systems fall below unity. Uncertainties in the cavity timescales that go into the calculation of the cavity power and the time required for the fading of star formation after cooling has been quenched would blur the division between red and blue systems. 

\section{Conclusions}
We have measured broadband optical colors and derived X-ray properties for a sample of CDGs in both cooling-flow and non-cooling-flow clusters to investigate possible connections between the presence of star formation and the properties of the cooling flow and the AGN. We show that, on similar spatial scales, the presence of central blue colors, indicative of active star formation, depends critically upon the presence of cooling gas with short cooling times. Blue cores are found to occur only in the clusters in our sample with central ICM cooling times below a threshold of $\sim 5 \times 10^8$ yr, with central entropies below $\sim 30$ kev cm$^{-2}$, and where the separation between the X-ray core and the CDG is less than $\sim 20$ kpc.  Finally, we find evidence that suggests that the lack of observable signatures of star formation in some systems with short cooling times could be due to excess heating by the central AGN that prevents net cooling from occurring.

Our results provide compelling evidence for the cycle of cooling, star formation, and energetic feedback. This cycle may be closely related to the physics responsible for several unsolved problems in galaxy formation, including the turnover at the bright end of the galaxy luminosity function and the preponderance of giant red galaxies, rather that giant blue galaxies at late times \citep{bowe06,crot06}. 

\acknowledgments
We thank Laura B\^{\i}rzan and Mark Voit for helpful discussions and Herald Ebeling for bringing the MACS sample to our attention. We also thank the referee for helpful comments. This research was funded in part by NASA Long Term Space Astrophysics Grant NAG4-11025 and by a generous grant from the Natural Sciences and Engineering Research Council of Canada. PEJN acknowledges NASA grant NAS8-01130.

\bibliography{master_references}

\end{document}